\documentclass[twocolumn,numberedappendix,twocolappendix,revtex4,iop]{openjournal}
\usepackage{graphicx,amsmath,amssymb,amstext}
\usepackage{amsbsy,amsfonts,amsthm,color}
\usepackage{times}
\usepackage{orcidlink}
\usepackage{hyperref}
\hypersetup{colorlinks=true, linkcolor=blue, citecolor=blue, filecolor=blue, urlcolor=blue}
\usepackage[utf8]{inputenc}
\usepackage{float}
\usepackage{xcolor}
\usepackage[T1]{fontenc}
\usepackage[title]{appendix}
\usepackage{lineno}
\usepackage{booktabs}
\usepackage{multirow}

\usepackage[font=footnotesize, labelfont=bf, labelsep=period]{caption}
\newcommand{\lcdm}{$\Lambda$CDM}
\newcommand{\Om}{\Omega_\mathrm{m}}
\newcommand{\threex}{$3\times2$-pt}
\newcommand{\vek}[1]{\mbox{\boldmath $#1$}}

\renewcommand{\arraystretch}{1.2}

\begin{document}

\title{DESI-DR1 $3 \times 2$-pt analysis: Consistent cosmology across weak lensing surveys \vspace{-4em}}
\shorttitle{DESI-DR1 $3\times2$-pt Analysis}

\author{
\parbox{\textwidth}{
A.~Porredon\orcidlink{0000-0002-2762-2024},$^{1,2,3,4}$
C.~Blake\orcidlink{0000-0002-5423-5919},$^{5}$
J.~U.~Lange\orcidlink{0000-0002-2450-1366},$^{6}$
N.~Emas,$^{5}$
J.~Aguilar,$^{7}$
S.~Ahlen\orcidlink{0000-0001-6098-7247},$^{8}$
A.~Bera,$^{9}$
D.~Bianchi\orcidlink{0000-0001-9712-0006},$^{10,11}$
D.~Brooks,$^{12}$
F.~J.~Castander\orcidlink{0000-0001-7316-4573},$^{13,14}$
T.~Claybaugh,$^{7}$
J.~Coloma Nadal,$^{15}$
A.~Cuceu\orcidlink{0000-0002-2169-0595},$^{7}$
K.~S.~Dawson\orcidlink{0000-0002-0553-3805},$^{16}$
A.~de la Macorra\orcidlink{0000-0002-1769-1640},$^{17}$
Biprateep~Dey\orcidlink{0000-0002-5665-7912},$^{18,19}$
P.~Doel,$^{12}$
A.~Elliott\orcidlink{0000-0001-6537-6453},$^{20,4}$
S.~Ferraro\orcidlink{0000-0003-4992-7854},$^{7,21}$
A.~Font-Ribera\orcidlink{0000-0002-3033-7312},$^{22}$
J.~E.~Forero-Romero\orcidlink{0000-0002-2890-3725},$^{23,24}$
C.~Garcia-Quintero\orcidlink{0000-0003-1481-4294},$^{25,26}$
E.~Gaztañaga\orcidlink{0000-0001-9632-0815},$^{13,27,14}$
S.~Gontcho A Gontcho\orcidlink{0000-0003-3142-233X},$^{7,28}$
G.~Gutierrez,$^{29}$
J.~Guy\orcidlink{0000-0001-9822-6793},$^{7}$
B.~Hadzhiyska\orcidlink{0000-0002-2312-3121},$^{30,21}$
H.~K.~Herrera-Alcantar\orcidlink{0000-0002-9136-9609},$^{31,32}$
S.~Heydenreich\orcidlink{0000-0002-7273-4076},$^{33}$
K.~Honscheid\orcidlink{0000-0002-6550-2023},$^{34,20,4}$
C.~Howlett\orcidlink{0000-0002-1081-9410},$^{35}$
D.~Huterer\orcidlink{0000-0001-6558-0112},$^{36,37}$
M.~Ishak\orcidlink{0000-0002-6024-466X},$^{9}$
S.~Joudaki\orcidlink{0000-0001-8820-673X},$^{1}$
R.~Joyce\orcidlink{0000-0003-0201-5241},$^{38}$
D.~Kirkby\orcidlink{0000-0002-8828-5463},$^{39}$
A.~Kremin\orcidlink{0000-0001-6356-7424},$^{7}$
A.~Krolewski,$^{40,41,42}$
O.~Lahav,$^{12}$
C.~Lamman\orcidlink{0000-0002-6731-9329},$^{4}$
M.~Landriau\orcidlink{0000-0003-1838-8528},$^{7}$
L.~Le~Guillou\orcidlink{0000-0001-7178-8868},$^{43}$
A.~Leauthaud\orcidlink{0000-0002-3677-3617},$^{33,44}$
M.~E.~Levi\orcidlink{0000-0003-1887-1018},$^{7}$
M.~Manera\orcidlink{0000-0003-4962-8934},$^{45,22}$
A.~Meisner\orcidlink{0000-0002-1125-7384},$^{38}$
R.~Miquel,$^{46,22}$
S.~Nadathur\orcidlink{0000-0001-9070-3102},$^{27}$
J.~ A.~Newman\orcidlink{0000-0001-8684-2222},$^{19}$
G.~Niz\orcidlink{0000-0002-1544-8946},$^{47,48}$
N.~Palanque-Delabrouille\orcidlink{0000-0003-3188-784X},$^{32,7}$
W.~J.~Percival\orcidlink{0000-0002-0644-5727},$^{40,41,42}$
C.~Poppett,$^{7,49,21}$
F.~Prada\orcidlink{0000-0001-7145-8674},$^{50}$
I.~P\'erez-R\`afols\orcidlink{0000-0001-6979-0125},$^{51}$
A.~Robertson,$^{38}$
G.~Rossi,$^{52}$
R.~Ruggeri\orcidlink{0000-0002-0394-0896},$^{53}$
E.~Sanchez\orcidlink{0000-0002-9646-8198},$^{1}$
C.~Saulder\orcidlink{0000-0002-0408-5633},$^{54}$
D.~Schlegel,$^{7}$
M.~Schubnell,$^{36,37}$
A.~Semenaite,$^{5}$
H.~Seo\orcidlink{0000-0002-6588-3508},$^{55}$
J.~Silber\orcidlink{0000-0002-3461-0320},$^{7}$
A.~Souki,$^{1}$
D.~Sprayberry,$^{38}$
G.~Tarl\'{e}\orcidlink{0000-0003-1704-0781},$^{37}$
M.~Vargas-Maga\~na\orcidlink{0000-0003-3841-1836},$^{17}$
B.~A.~Weaver,$^{38}$
C.~Zhou,$^{33}$
R.~Zhou\orcidlink{0000-0001-5381-4372},$^{7}$
and H.~Zou\orcidlink{0000-0002-6684-3997}$^{56}$
}
\begin{center} (DESI Collaboration) \end{center}
\vspace{0.2em}
{\it (Affiliations can be found after the references)}
}

\thanks{$^*$E-mail: annamaria.porredon@ciemat.es}
\shortauthors{A.~Porredon et al.}

\label{firstpage}

\begin{abstract}
We present a joint cosmological analysis of projected galaxy clustering observations from the Dark Energy Spectroscopic Instrument Data Release 1 (DESI-DR1), and overlapping weak gravitational lensing observations from three datasets: the Kilo-Degree Survey (KiDS-1000), the Dark Energy Survey (DES-Y3), and the Hyper-Suprime-Cam Survey (HSC-Y3).  This combination of large-scale structure probes allows us to measure a set of $3 \times 2$-pt correlation functions, breaking the degeneracies between parameters in cosmological fits to individual observables.  We obtain mutually-consistent constraints on the parameter $S_8 = \sigma_8 \sqrt{\Omega_{\rm m}/0.3} = 0.786^{+0.022}_{-0.019}$ from the combination of DESI-DR1 and DES-Y3, $S_8 = 0.760^{+0.020}_{-0.018}$ from KiDS-1000, and $S_8 = 0.771^{+0.026}_{-0.027}$ from HSC-Y3.  These parameter determinations are consistent with fits to the \textit{Planck} Cosmic Microwave Background dataset, albeit with $1.5-2\sigma$ lower values in the $S_8-\Om$ plane.  We perform our analysis with a unified pipeline tailored to the requirements of each cosmic shear survey, which self-consistently determines cosmological and astrophysical parameters.  We generate an analytical covariance matrix for the correlation data including all cross-covariances between probes, and we design a new blinding procedure to safeguard our analysis against confirmation bias, whilst leaving goodness-of-fit statistics unchanged. Our study is part of a suite of papers that present joint cosmological analyses of DESI-DR1 and weak gravitational lensing datasets.
\end{abstract}
\maketitle

\section{Introduction}
\label{sec:intro}

A central goal of modern cosmology is to define and scrutinise the model that describes the composition, initial conditions and evolution of the Universe.  Given a set of compelling unsolved problems, including the unknown physics represented by dark matter and dark energy, we seek new information that might refine our knowledge \citep[for reviews, see][]{2013PhR...530...87W, 2019LRR....22....1I, 2022LRR....25....6M, 2023A&ARv..31....2H}.  As cosmological datasets continue to expand impressively, an important theme is to combine these datasets in unified analysis frameworks.  Such joint analyses serve to improve the statistical precision of our cosmological parameter determinations, mitigate against key systematic errors, and stress-test the self-consistency of our model framework against the baseline established by analyses of the Cosmic Microwave Background (CMB) radiation \citep{Planckcosmo}.  Such comparisons have already revealed a variety of ``tensions'' and ``intriguing hints'' which guide future exploration: including disagreements between determinations of the local expansion rate \citep{2016ApJ...826...56R, 2021CQGra..38o3001D, 2021ApJ...919...16F, 2022ApJ...934L...7R, 2023Univ....9..393V}, debates over the amplitude of matter fluctuations \citep{2017MNRAS.467.3024L, 2018MNRAS.474.4894J, 2021A&A...646A.140H}, and suggestions that the physics of dark energy might evolve with time \citep{2025PhRvD.112h3514A, 2025PhRvD.112h3515A, 2025JCAP...02..021A}.

The large-scale structure of the Universe as mapped by spectroscopic redshift surveys, and the gravitational lensing of light by this structure as probed by photometric imaging surveys, have long been seen as highly complementary cosmological probes \citep[e.g.,][]{2012MNRAS.422.1045C, 2012MNRAS.422.2904G, 2013arXiv1308.6070D, 2015MNRAS.451.4424K}.  A set of $3 \times 2$-pt correlation statistics can be constructed from these datasets: cosmic shear, the correlated distortions in galaxy shapes induced by weak gravitational lensing by the cosmic web; galaxy clustering, the correlations between galaxy positions caused by gravitational physics; and galaxy-galaxy lensing, the tangential shear of background source galaxies around foreground lens galaxies.  The amplitude and scale-dependence of these correlations may be accurately predicted by lensing and clustering theory, allowing us to infer the values of the cosmological parameters which control the expansion, growth and distribution of matter.  However, correlations are also induced by a variety of astrophysical effects, including intrinsic alignments, unknown photometric redshift distributions, baryonic effects in the matter power spectrum, shape measurement calibrations, and the biasing of galaxies with respect to matter fluctuations.  Testing the cosmological model in light of these potential complications requires us to create a joint analysis framework which can be applied across survey datasets. 

In this paper, we present a unified cosmology analysis of $3 \times 2$-pt statistics utilising the large-scale structure dataset assembled by the Dark Energy Spectroscopic Instrument (DESI) in conjunction with three important weak lensing (WL) surveys.  We hence use DESI galaxies to form the lens sample for the $3 \times 2$-pt correlations, instead of photometric galaxies from the imaging surveys.  DESI, which has been operating its main survey program at Kitt Peak National Observatory since 2021, has already assembled the largest existing spectroscopic redshift dataset for cosmology, exceeding previous programs by more than an order of magnitude \citep{2016arXiv161100036D, 2016arXiv161100037D, 2024AJ....167...62D}.  Our study is based on the first substantial assembly of DESI data, compiled from the first year of operations \citep[Data Release 1,][]{2025arXiv250314745D}.  Together with DESI data, we also use publicly-available weak lensing catalogues from the Dark Energy Survey Year 3 \citep[DES-Y3,][]{2021ApJS..254...24S, 2021MNRAS.504.4312G}, the Kilo-Degree Survey \citep[KiDS-1000,][]{2019A&A...625A...2K, 2021A&A...645A.105G}, and the Hyper Suprime-Cam survey Year 3 \citep[HSC-Y3,][]{2022PASJ...74..247A, 2022PASJ...74..421L}.  Our work builds upon a wealth of cosmological studies already presented by these collaborations \citep[e.g.,][]{2017MNRAS.465.1454H, 2018PhRvD..98d3526A, 2019PASJ...71...43H, 2020PASJ...72...16H, 2020A&A...633A..69H, 2021A&A...646A.140H, 2022PhRvD.105b3520A, 2023PhRvD.108l3518L, 2023PhRvD.108l3519D, KiDSLegacy_CS_Wright2025}.  In particular, \cite{2023OJAp....6E..36D, 2024JCAP...08..024G, 2025OJAp....8E.139J} have previously performed analyses under a unified modelling pipeline.

Our work presents several key advances in the $3 \times 2$-pt cosmology program.  First, DESI-DR1 provides an unprecedented set of new spectroscopic lenses, overlapping with all the weak lensing surveys considered in our study and hence allowing the measurement of consistent galaxy-galaxy lensing cross-correlations.  Second, spectroscopic clustering samples allow the usage of the projected correlation function $w_p(r_p)$ for the lens auto-correlations, rather than the angular correlations $w(\theta)$ that are accessible in photometric datasets.  The projected correlation function can be measured with higher signal-to-noise ratio than $w(\theta)$, since the signal is integrated over a narrower radial extent, and hence experiences less dilution.  As a third aspect, our work has extended $3 \times 2$-pt analytical covariance frameworks to include $w_p(r_p)$ rather than $w(\theta)$.  Finally, we have created a new unified analysis pipeline, tailored to the requirements of each cosmic shear survey, which self-consistently determines cosmological and astrophysical parameters.

The current paper is the culmination of a set of preparatory studies which have tested and validated the different components of our analysis.  \cite{2025OJAp....8E..24B} presented a ``mock challenge'' study of simulated catalogues representing our key datasets.  \cite{2024MNRAS.533..589Y} tested our selection of lens galaxies from the parent DESI dataset and the associated covariance for galaxy-galaxy lensing, and \cite{2025arXiv250621677H} presented galaxy-galaxy measurements for our weak lensing survey sources and tested their consistency.  \cite{Nimas2025} performed validation of the scale cuts used in this study, and used shear ratios to test for the presence of systematic effects in lensing amplitudes, and \cite{Ruggeri2025} analysed the position cross-correlations between the lensing and DESI catalogues to validate the source redshift distributions via the clustering-redshift technique.  As companion papers to our work, \cite{Semenaite2025} and \cite{Joe_in_prep} present joint cosmological analyses of lensing and clustering multipoles which include the effect of redshift-space distortions, in contrast to the projected clustering statistic adopted in our study, and \cite{Lange2025} present full-scale analyses of galaxy-galaxy lensing and clustering statistics using emulators calibrated by simulations, which permit models to be extended to smaller scales than utilised in our study.

Our paper is structured as follows: in Sec.~\ref{sec:data} we overview the large-scale structure and weak lensing datasets used in this study.  In Sec.~\ref{sec:methodology} we describe our methodology for modelling and measuring the $3 \times 2$-pt correlations, for determining the covariance of the statistics, and for applying the data-vector blinding we implement to mitigate against confirmation bias.  In Sec.~\ref{sec:results} we describe our cosmological results and robustness tests, and in Sec.~\ref{sec:discussion} we compare our results with other studies.  We conclude our work in Sec.~\ref{sec:concl}.

\section{Data}
\label{sec:data}

\subsection{DESI Data Release 1}

The Dark Energy Spectroscopic Instrument (DESI) is a multi-object spectrograph installed at the 4-metre Mayall Telescope at Kitt Peak National Observatory.  Over its 8-year main survey program, which commenced in May 2021, DESI will collect over 60 million spectra of galaxies and quasars across $17{,}000$\,deg$^2$ \citep{2016arXiv161100036D, 2016arXiv161100037D, 2024AJ....167...62D}, increasing the size of existing large-scale structure samples by more than an order-of-magnitude.  The instrument design is described by \cite{2022AJ....164..207D}, including an optical corrector \citep{2024AJ....168...95M} and a robotic positioner \citep{2023AJ....165....9S} that can assign up to 5000 fibres to targets in the focal plane \citep{2024AJ....168..245P}, where the resulting data are processed by the DESI spectroscopic pipeline \citep{2023AJ....165..144G}.  In the DESI observing strategy, described by \cite{2023AJ....166..259S}, the survey obtains spectra for four principal target classes, photometrically-selected from the DESI Legacy optical imaging surveys \citep{2019AJ....157..168D}: these components are called the Bright Galaxy Survey \citep[BGS,][]{2023AJ....165..253H}, Luminous Red Galaxy Survey \citep[LRG,][]{2023AJ....165...58Z}, Emission Line Galaxy Survey \citep[ELG,][]{2023AJ....165..126R}, and Quasar Survey \citep[QSO,][]{2023ApJ...944..107C}.

We use DESI Data Release 1 \citep[DR1,][]{2025arXiv250314745D} as the dataset for the galaxy-galaxy lensing and galaxy clustering measurements in this analysis.  DESI-DR1 consists of all data acquired during the first 13 months of the main survey up to June 2022, including high-confidence redshifts for 13.1M galaxies \citep{2025arXiv250314745D}.  In our analysis we use catalogues from the Bright Galaxy Survey \citep{2023AJ....165..253H} and Luminous Red Galaxy survey \citep{2023AJ....165...58Z}.  Bright Galaxies span redshift range $0.1 < z < 0.4$, which we divide into three narrower redshift ranges $(0.1-0.2, 0.2-0.3, 0.3-0.4)$.  Within each of these BGS redshift bins, we apply luminosity cuts to select lens samples \citep{2024MNRAS.533..589Y}, which we use for galaxy-galaxy lensing and clustering measurements \citep{2025arXiv250621677H}.  In our study, we hence select fainter BGS samples than utilised for the Key Project Baryon Acoustic Oscillation and full shape analyses \citep{2025JCAP...07..017A, desi_fs}, to increase the signal-to-noise of the resulting galaxy-galaxy lensing measurements \citep{2024MNRAS.533..589Y}.  To these three BGS lens samples, we add LRG catalogues in redshift ranges $(0.4-0.6, 0.6-0.8)$, which are the same samples as used in the DESI Key Project.  We do not use the $0.8 < z < 1.1$ LRG sample in our study, because of its poor efficiency for lensing with current weak lensing surveys (and we also do not utilise the higher-redshift DESI Emission Line Galaxy and Quasar samples for the same reason).

\subsection{DES Year 3}

In addition to the DESI spectroscopic large-scale structure dataset, we employ weak gravitational lensing shear measurements from three current photometric imaging surveys.  First, we include the Dark Energy Survey Year 3 (DES-Y3) shear catalogue \citep{2021MNRAS.504.4312G} in our analysis.  The Dark Energy Survey has used the Dark Energy Camera mounted on the Blanco 4-metre Telescope \citep{2015AJ....150..150F} to conduct the largest existing weak lensing survey.  The DES-Y3 catalogue, which contains more than $10^8$ objects over $4{,}143$ deg$^2$, was based on imaging data from the first three years of DES operation between 2013 and 2016 \citep{2021ApJS..254...24S}.  The shear catalogue was created from this imaging dataset using the self-calibrating shear measurement pipeline \textsc{metacalibration} \citep{2017ApJ...841...24S}, and further re-calibrated for shear bias and blending by a suite of image simulations \citep{2022MNRAS.509.3371M}, and has a weighted source number density of 5.6 arcmin$^{-2}$ and corresponding shape noise $\sigma_e = 0.26$.  We analysed DES-Y3 sources in the four tomographic bins created by the collaboration \citep{2021MNRAS.505.4249M}, split by photometric redshifts $[0.00, 0.36, 0.63, 0.89, 2.00]$.  The DES collaboration has performed validations and cosmological fits based on these catalogues \citep[see,][]{2022PhRvD.105b3520A, 2022PhRvD.105b3515S, 2022PhRvD.105b3514A}.  We note that, in our analysis, we use a revised version of the DES-Y3 catalogue with updated source tomographic bin assignments, matching the current files available in the DES-Y3 data release page\footnote{\url{https://des.ncsa.illinois.edu/releases/y3a2/Y3key-catalogs}} (see also footnote 5 of \citealt{DESY3blueshear-2024}). The DES-Y3 catalogue has an overlap of 851.3 deg$^2$ with the DESI-DR1 footprint.

\subsection{KiDS-1000}

The second weak lensing dataset we include in our analysis is the Kilo-Degree Survey (KiDS-1000) weak lensing catalogue \citep{2021A&A...645A.105G}, which is part of the fourth public KiDS data release \citep{2019A&A...625A...2K}.  KiDS has utilised the OmegaCAM instrument on the Very Large Telescope Survey Telescope (VST) at the Paranal Observatory to image the sky in optical filters $u, g, r, i$.  Complementary imaging in the near-infrared bands $Z, Y, J, H, K_s$ has been obtained by the VISTA-VIKING survey, resulting in a deep, wide, nine-band imaging dataset with improved photometric redshift calibration \citep{2021A&A...647A.124H}.  KiDS-1000 includes $1{,}006$ deg$^2$ of lensing data, consisting of 21 million galaxies with an effective number density 6.2 arcmin$^{-2}$, with source shapes measured using the {\it lens}fit model fitting technique \citep{2013MNRAS.429.2858M, 2017MNRAS.467.1627F}.  Our analysis uses the five tomographic source samples defined by KiDS, which are divided by photometric redshifts $[0.1, 0.3, 0.5, 0.7, 0.9, 1.2]$, where the redshift distribution of the sources is calibrated as described by \cite{2021A&A...647A.124H}.  Cosmological fits to the KiDS-1000 dataset were presented by \cite{2021A&A...646A.140H}, and joint cosmological analyses of DES-Y3 and KiDS-1000 have been performed by \cite{2023OJAp....6E..36D}.  The area of overlap of KiDS-1000 and DESI-DR1 is 446.8 deg$^2$.  Whilst our work was in progress, cosmological analysis of the extended KiDS-Legacy survey was presented by that collaboration \citep{KiDSLegacy_CS_Wright2025}.

\subsection{HSC Year 3}

The third and final weak lensing dataset we include in our study is the Hyper Suprime-Cam survey Year 3 catalogue \citep[HSC-Y3][]{2022PASJ...74..421L}.  The HSC survey is using the 8.2-metre Subaru Telescope to create the deepest large-scale weak lensing dataset currently available, imaging several regions of sky in $g, r, i, z, Y$ filters, where the $i$-band seeing is around $0.6\arcsec$ \citep{2022PASJ...74..247A}.  The HSC-Y3 catalogue is based on data acquired between 2014 and 2019, covering 434 deg$^2$ of sky with effective source density 19.9 arcmin$^{-2}$.  The catalogue is based on source shape measurements performed using a re-Gaussianization PSF correction method \citep{2018PASJ...70S..25M}, further calibrated by realistic image simulations that emulate survey observation conditions \citep{2022PASJ...74..421L}.  Our work uses the four tomographic source samples defined by the HSC collaboration, with photometric redshift divisions $[0.3, 0.6, 0.9, 1.2, 1.5]$, and redshift distributions inferred by \cite{2023MNRAS.524.5109R}.  Cosmological fits to these datasets have been presented by \cite{2023PhRvD.108l3518L} and \cite{2023PhRvD.108l3519D}. We note that, recently, \cite{Choppin2025,Choppin2025b} produced an updated calibration of the HSC-Y3 redshift distributions using spectroscopic data from DESI DR1 and DR2 that was not available in \cite{2023MNRAS.524.5109R}. In this work, we use the original redshift distributions from \cite{2023MNRAS.524.5109R} and the priors from the fiducial HSC-Y3 analyses \citep{2023PhRvD.108l3518L,2023PhRvD.108l3519D} (see Sec.~\ref{subsec: params} for details). The HSC-Y3 dataset has almost complete overlap with DESI-DR1.  Sky coverage maps of the different datasets used in our analysis can be found in Fig.~1 of \cite{Nimas2025}.

\section{Methodology}
\label{sec:methodology}

In this section we describe our methodology for inferring cosmological parameters from the combination of DESI-DR1 galaxy clustering data and DES, KiDS, and HSC weak lensing data. We assume two main cosmological models: flat \lcdm\, and flat $w$CDM, with $w$ being the equation of state for dark energy, which we assume is constant in time. 

We fit these cosmological models to a combination of $3 \times 2$-pt configuration-space correlation measurements generated from the galaxy source and lens samples as described in Sec.~\ref{subsec:measurements}.  We use the cosmic shear correlation functions for the tomographic source sub-samples of the weak lensing surveys, $\xi_{\pm}(\theta)$, as a function of angular separation $\theta$; the projected galaxy correlation functions of each DESI lens sample, $w_p(r_p)$, as a function of projected separation $r_p$; and the average tangential shear of the source samples around the lens galaxies, $\gamma_t(\theta)$.  For the lens clustering, we note that we utilise the projected correlation function in preference to the angular correlation function, $w(\theta)$, since $w_p(r_p)$ can be measured with a higher signal-to-noise ratio (S/N) than $w(\theta)$ for spectroscopic lens samples. \cite{2025arXiv250621677H} also provides measurements of the excess surface mass density $\Delta \Sigma (r_p)$ for galaxy-galaxy lensing. Nevertheless, we opt to stick with $\gamma_t(\theta)$ because the S/N gain is not as clear in this case \citep{Shirasaki2018}, and this allows us to use the modelling framework from \cite{2022PhRvD.105b3520A}. 

\subsection{Modelling}
\label{subsec:model}

\subsubsection{Cosmic shear}
\label{subsec:shear}

We follow the DES-Y3 framework described in \cite{Krause2021} and \cite{2022PhRvD.105b3520A} to model the angular two-point correlation functions of cosmic shear and galaxy-galaxy lensing.
For cosmic shear, and for two given redshift bins $i$ and $j$, these functions can be expressed in terms of the convergence angular power spectrum $C_{\kappa}(\ell)$ at an angular wavenumber $\ell$:
\begin{equation}
\begin{aligned}
    \xi_{\pm}^{ij}(\theta) = \sum_{\ell} \frac{2\ell + 1}{2\pi\ell^2(\ell + 1)^2} [G_{\ell,2}^{+}(\cos \theta) \pm G_{\ell,2}^{-}(\cos \theta)]\\
    \times \left[ C_{\kappa, \mathrm{EE}}^{ij}(\ell) \pm C_{\kappa, \mathrm{BB}}^{ij}(\ell)\right],
\end{aligned}
\end{equation}
where $G_{\ell}^{\pm}$ are computed from the Legendre polynomials $P_{\ell}$ (see Eq.~(4.19) of \cite{Stebbins1996} for their full expression). The two-point correlation functions are then averaged over each angular bin $[\theta_{\min}, \theta_{\max}]$ following the expressions from \cite{Fang_2020}. 
The observed shear angular power spectra can be decomposed into E and B modes, with the B-mode contribution $C_{\kappa, \mathrm{BB}}^{ij}(\ell)$ originating from the intrinsic alignments of galaxies (see e.g.\ \citealt{2015PhR...558....1T} and \citealt{IAreview2024} for reviews):
\begin{align}
 \label{eq:C_ells_lensing}
    C_{\kappa, \mathrm{EE}}^{ij}(\ell) = C_{\kappa\kappa}^{ij}(\ell)+C_{\kappa \rm I_E}^{ij}(\ell) + C_{\kappa \rm I_E}^{ji}(\ell) + C_{\rm I_E I_E}^{ij}(\ell), \\
    C_{\kappa, \mathrm{BB}}^{ij}(\ell) = C_{\rm I_B I_B}^{ij}(\ell), 
\end{align}
where $\rm{I}$ denotes an intrinsic alignment contribution.

The angular convergence power spectrum $C_{\kappa\kappa}(\ell)$ can be obtained from the non-linear 3D matter power spectrum $P_{\mathrm{mm}}$. Under the Limber approximation \citep{Limber1953,LoVerde2008} and assuming a flat spatial geometry,
\begin{equation}
C_{\kappa \kappa}^{ij}(\ell) = \int_0^{\chi_{\rm H}}\mathrm{d} \chi \frac{W_{\kappa}^i(\chi)W_{\kappa}^j(\chi)}{\chi^2}P_{\mathrm{mm}}\left( \chi, k=\dfrac{\ell+1/2}{z(\chi)}\right),    
\end{equation}
where $\chi_{\rm H}$ is the horizon distance, $z(\chi)$ is the redshift at the comoving distance $\chi$, and $W_{\kappa}^i(\chi)$ is the lensing efficiency kernel for the redshift bin $i$,
\begin{equation}
W_{\kappa}^i(\chi) = \frac{3H_0^2\Om}{2c^2}\frac{\chi}{a(\chi)}\int_{\chi}^{\chi_{\rm H}}\mathrm{d}\chi'n_{\rm s}^i(\chi')\frac{\chi'-\chi}{\chi'},
\end{equation}
with $c$ being the speed of light, $a$ the scale factor, $H_0$ the Hubble constant, $\Om$ the matter energy density, and  $n_{\rm s}^i$ the normalized redshift distribution of source galaxies in bin $i$.
We calculate the matter power spectrum $P_{\rm mm}$ from the \textsc{CAMB} \citep{camb} implementation in \textsc{CosmoSIS}\footnote{\url{https://cosmosis.readthedocs.io}} \citep{cosmosis}, and we use \textsc{HMcode2020} with baryonic feedback \citep{hmcode2020} as our baseline model for the non-linear matter power spectrum.

For the intrinsic alignments, we assume the non-linear linear-alignment (NLA) model \citep{NLA_Bridle_2007} as fiducial, since it has been widely used in previous weak lensing analyses \citep[see, e.g.,][]{2021A&A...645A.104A, 2022PhRvD.105b3520A, HSC3x2_Miyatake2023, KiDSLegacy_CS_Wright2025, DECADE_cs_2025} and is the fiducial choice in the joint cosmological analysis of DES-Y3 and KiDS-1000 cosmic shear \citep{2023OJAp....6E..36D}. This model describes the linear tidal alignment of galaxies with the density field \citep{HirataIA_2004} and includes an ``ad hoc'' non-linear correction to the linear matter power spectrum, as motivated by intrinsic alignment measurements in the Sloan Digital Sky Survey \citep{NLA_Hirata_2007}. The NLA intrinsic alignment power spectra $C_{\kappa \rm I_E}(\ell)$ and $C_{\rm I_E I_E}(\ell)$  from Eq.~\eqref{eq:C_ells_lensing} are described in Eqs.~(3-5) in \cite{NLA_Bridle_2007}, while $C_{\rm I_B I_B}(\ell) = 0$. With this model, the convergence-intrinsic and intrinsic-intrinsic power spectra are related to the nonlinear matter power spectrum through a redshift-dependent amplitude,
\begin{align}
P_{\kappa \rm I_E} = c(z) P_{\rm mm}, \quad P_{\rm I_E I_E}= c^2(z)P_{\rm mm}, \\
c(z) = - A_{\rm IA} \frac{\bar{C}\rho_{\rm c}\Om}{D(z)}\left( \frac{1+z}{1+z_0}\right)^{\eta_{\rm IA}},
\end{align}
where $D(z)$ is the growth function, $\rho_{\rm c}$ is the critical density, $z_0=0.62$ is the pivot redshift, and $\bar{C}=5\times 10^{-14}M_{\odot}h^{-2}\mathrm{Mpc}^2$ is obtained from SuperCOSMOS \citep{supercosmos}. $A_{\rm IA}$ and $\eta_{\rm IA}$ are free parameters that we vary in our analysis (see Sec.~\ref{subsec: params} for more details).

In some cases, we also consider the Tidal Alignment and Tidal Torquing model (TATT, \citealt{TATT}), which extends the linear alignment model with the tidal torquing alignment mechanism. With this model, we have five free nuisance parameters: the tidal alignment amplitude, $A_1$, with redshift evolution $\eta_1$; the tidal torquing amplitude, $A_2$, with redshift evolution $\eta_2$, and a linear bias amplitude $b_{\rm TA}$. See Eqs.~(37-39) in \citealt{TATT} for the expressions of the TATT intrinsic alignment power spectra, including a non-zero B-mode contribution $C_{\rm I_B I_B}(\ell)$. In the limit where $A_2, \eta_2, b_{\rm TA} \rightarrow 0$, the TATT model reduces to the NLA model, where $A_1$ and $\eta_1$ correspond to $A_{\rm IA}$ and $\eta_{\rm IA}$, respectively.

\subsubsection{Galaxy-galaxy lensing}
\label{subsec:ggl}

Similarly to cosmic shear, the galaxy-galaxy lensing two-point correlation function $\gamma_{\rm t}^{ij}$ for lens bin $i$ and source bin $j$ can be expressed in terms of the density-convergence angular power spectrum $C_{\delta_{\rm g},\kappa}^{ij}(\ell)$ as,
\begin{equation}
    \gamma_{\rm t}^{ij} (\theta) = \sum_{\ell}\frac{2\ell + 1}{4\pi \ell (\ell + 1)} \, P_{\ell}^2(\cos \theta) \, C_{\delta_{\rm g}\kappa_{\rm tot}}^{ij}(\ell),
\end{equation}
with $P_{\ell}^2$ being the Legendre polynomials, which are averaged over each angular bin in the same way as $\xi_{\pm}$, following \cite{Fang_2020}.
The observed density-convergence angular power spectrum also includes cross-terms related to intrinsic alignments ($\rm I_E$) and lens magnification ($\delta_{\rm mag}$),
\begin{equation}
   C_{\delta_{\rm g},\kappa_{\rm tot}}^{ij}(\ell) = C_{\delta_{\rm g}\kappa}^{ij}(\ell) + C_{\delta_{\rm g}\rm I_E}^{ij}(\ell) + C_{\delta_{\rm mag},\kappa}^{ij}(\ell) + C_{\delta_{\rm mag},\rm I_E}^{ij}(\ell).
\end{equation}
Under the Limber approximation and assuming a flat spatial geometry we obtain,
\begin{equation}
  C_{\delta_{\rm g}\kappa}^{ij}(\ell) = \int_0^{\chi_{\rm H}}\mathrm{d} \chi \frac{W_{\delta_{\rm g}}^i(\chi)W_{\kappa}^j(\chi)}{\chi^2}P_{\mathrm{mm}}\left( \chi, k=\dfrac{\ell+1/2}{z(\chi)}\right),   
\end{equation}
with $W_{\delta_{\rm g}}^i(\chi)$ being the density kernel in lens redshift bin $i$,
\begin{equation}
    W_{\delta_{\rm g}}^i(\chi) = b^i(z(\chi)) \, n^i_{\rm l}(z(\chi)) \, \frac{\mathrm{d z}}{\mathrm{d}\chi},
\end{equation}
where $b^i$ is the galaxy bias and $n^i_{\rm l}(\chi)$ is the normalised redshift distribution of galaxies in lens bin $i$. In our analysis, we assume a linear galaxy bias model with one free parameter $b^i$ per lens redshift bin $i$. 

Lens magnification is an effect produced by gravitational lensing that impacts the number of galaxies observed in a given area of the sky. For galaxy-galaxy lensing, this effect is particularly important when the lens galaxies are behind the sources (see, e.g., \citealt{Cross_2024_IGGL} and \citealt*{Elvin-Poole2023}). The effect of magnification on the projected number density contrast at position $\mathbf{\hat{n}}$ and redshift bin $i$ can be expressed in terms of the convergence $\kappa^i(\mathbf{\hat{n}})$ experienced by lens galaxies in bin $i$,
\begin{equation}
    \delta_{\rm mag}^i(\mathbf{\hat{n}}) = C^i \, \kappa^i(\mathbf{\hat{n}}),
\end{equation}
where $C^i$ is a constant of proportionality given by the response of the number of selected objects per area to a change in $\kappa$. This constant can be obtained from the measured slope of the luminosity function $\alpha^i$ at the faint end of the lens galaxy distribution in redshift bin $i$ \citep{Joachimi2010}: $C^i=2(\alpha^i - 1)$. We assume for $\alpha^i$ the values provided for our lens redshift bins in Table~3 of \cite{2025arXiv250621677H}.

Given that the tangential shear two-point correlation function $\gamma_{\rm t}$ is a non-local measure of the galaxy-mass cross-correlation function, we follow \cite{MacCrann2020}  and analytically marginalise over the mass enclosed below the angular scales included in the analysis. The analytical marginalisation is implemented by modifying the inverse of the covariance matrix when calculating the likelihood. We use the implementation and priors from the DES-Y3 analysis, which are described in detail in \cite{Pandey2022}.    

\subsubsection{Projected galaxy clustering}
\label{subsec:wp}

As mentioned, we take advantage of the spectroscopic data from DESI and use the projected galaxy clustering correlation function $w_p(r_p)$, instead of the angular correlation function $w(\theta)$. We first follow \cite{Kaiser1987_rsd} to obtain the redshift-space galaxy power spectrum from the matter power spectrum $P_{\rm mm}$, as a function of the Fourier modes $k$ and the cosine of the angle between the direction of the wavevector and the line of sight, $\mu$,
\begin{equation}
    P_{\rm gg}(k, \mu) = b^2 \, (1 + \beta\mu^2)^2 \, P_{\rm mm}(k),
\end{equation}
with $\beta\equiv f/b$, where $f$ is the logarithmic derivative of the growth factor $D(z)$ and $b$ is the linear bias factor. $P_{\rm gg}(k, \mu)$ can then be expanded into Legendre polynomials of order 0, 2, and 4,
\begin{equation}
\begin{gathered}
    P_{\rm gg,0} (k) = b^2\left(1 + \dfrac{2}{3}\beta +\dfrac{1}{5}\beta^2\right)P_{\rm mm}(k), \\
    P_{\rm gg,2} (k) = b^2\left(\dfrac{4}{3}\beta +\dfrac{4}{7}\beta^2\right)P_{\rm mm}(k), \\
    P_{\rm gg,4} (k) = b^2\left(\dfrac{8}{35}\beta^2\right)P_{\rm mm}(k).
\end{gathered}
\end{equation}
From these power spectrum multipoles, we can obtain the correlation function multipoles $\xi_{\ell}(s)$ through a Hankel transform \citep{Cole1994},
\begin{equation}
    \xi_{\rm gg,\ell}(s) = i^{\ell}\dfrac{1}{2\pi ^2}\int_0^{\infty} \mathrm{d}k \, k^2 \, P_{\rm gg,\ell}(k) \, j_{\ell}(ks),
    \label{eq:hankl}
\end{equation}
where $s$ is the separation between two galaxies and $j_{\ell}(ks)$ is the spherical Bessel function of order $\ell$. We can decompose $s$ into perpendicular, $r_p$, and parallel, $r_{\pi}$, components to the line of sight: $s=\sqrt{r_p^2+r_{\pi}^2}$. The projected correlation function $w_p(r_p)$ is then defined as the projection of $\xi_{\rm gg}(r_p, r_{\pi})$ along the line of sight component,
\begin{equation}
    w_p(r_p)=2\int_0^{r_{\pi,\max}} \mathrm{d}r_{\pi}  \, \xi_{\rm gg}(r_p, r_{\pi}),
\end{equation}
where we assume $r_{\pi,\max}=100\,h^{-1}\rm Mpc$ for the integration limit, matching the choice in the correlation measurement described below, and $\xi_{\rm gg}(r_p, r_{\pi})$ is obtained from the sum of the correlation function multipoles from Eq.~\eqref{eq:hankl}, weighted by the corresponding Legendre polynomials of order 0, 2 and 4.

We use the \textsc{hankl}\footnote{\url{https://hankl.readthedocs.io/}} package \citep{hankl} to evaluate the Hankel transform of Eq.~\eqref{eq:hankl} with the FFTlog algorithm. As in the previous subsections, we use \textsc{HMcode2020} for the non-linear matter power spectrum $P_{\rm mm}$, and assume a linear galaxy bias model with one free parameter $b^i$ per lens redshift bin $i$. 

\subsection{Measurements}
\label{subsec:measurements}

In this section we briefly describe how we obtain or perform each correlation measurement.  For the cosmic shear correlation functions between the tomographic source sub-samples of each weak lensing survey, we use the publicly-released measurements of the DES-Y3, KiDS-1000 and HSC-Y3 collaborations.  DES-Y3 has presented $\xi_{\pm}(\theta)$ measurements in 20 logarithmically-spaced separation bins in the range $2.5 < \theta < 250'$ \citep{2022PhRvD.105b3520A}; KiDS-1000 used 9 separation bins between limits $0.5 < \theta < 300'$ \citep{2021A&A...645A.104A}; and HSC-Y3 performed cosmic shear measurements using 7 angular bins in the range $7.1 < \theta < 56.6'$  for $\xi_+$, and 7 bins in the range $31.2 < \theta < 248'$ for $\xi_-$ \citep{2023PhRvD.108l3518L}.

Our measurements of the projected galaxy correlation functions of the five DESI-DR1 lens samples, within redshift ranges $(0.1-0.2, 0.2-0.3, 0.3-0.4, 0.4-0.6, 0.6-0.8)$, are described by \cite{2025arXiv250621677H}.  In brief, the correlation function measurements were performed using {\sc pycorr}\footnote{\url{https://py2pcf.readthedocs.io/}}, which wraps a modified version of the {\sc corrfunc} package\footnote{\url{https://corrfunc.readthedocs.io/}} \citep{2020MNRAS.491.3022S}.  We performed measurements in 15 logarithmic separation bins in the range $0.08 < r_p < 80 \, h^{-1}$Mpc, where we will later exclude small separations which present modelling difficulties (see Sec.~\ref{subsec: scale cuts}).  When constructing the projected correlation function, we sum over line-of-sight separations $|r_{\pi}| \le 100 \, h^{-1}$Mpc (choosing this limit to exceed the maximum value of $r_p$ adopted).  Our clustering measurements utilise accompanying DESI random catalogues \citep{2025JCAP...07..017A}, and employ weights which compensate for target selection systematics, spectroscopic incompleteness and fibre collisions.  However, we exclude small-scale PIP (pairwise inverse probability) weights \citep{2025JCAP...04..074B}, because their effect is negligible at the separations relevant for our study.  We also apply integral constraint corrections to the $w_p(r_p)$ measurements.

Finally, we measured the average tangential shear of all the tomographic source sub-samples around each DESI lens dataset.  Our galaxy-galaxy lensing measurements using these catalogues are again discussed by \cite{2025arXiv250621677H}.  We correct our lensing measurements for multiplicative shear bias and dilutions due to source photometric redshift errors, and we also subtract the signal around random lenses.  We performed measurements in 15 logarithmic angular separation bins in the range $0.3 < \theta < 300$ arcmin using the \textsc{dsigma} code\footnote{\url{https://dsigma.readthedocs.io/}} \citep{2022ascl.soft04006L}, where we will later discuss in Sec.~\ref{subsec: scale cuts} the scale cuts we apply in model fits.  Before performing the galaxy-galaxy lensing measurement, we cut the source and lens samples to overlapping sky areas \citep{2025arXiv250621677H}, whereas we use the full DESI sky coverage to perform the projected correlation function measurements described above.

\subsection{Parametrisation and priors}
\label{subsec: params}

\begin{table}
    \caption{Cosmological and astrophysical parameters, and their priors, used in the $\Lambda$CDM and $w$CDM analyses. The parameter $w$ is fixed to $-1$ in $\Lambda$CDM.  Square brackets denote a flat prior, while parentheses denote a Gaussian prior of the form $\mathcal{N}(\mu,\sigma)$.} 
    \label{tab:params}
	\centering	
	\begin{tabular}{ccc}
		\toprule
		Parameter & Fiducial &Prior\\
        \midrule
		\multicolumn{3}{c}{\textbf{Cosmology}} \\ [+0.1cm] 
		$\Omega_{\rm m}$ &  0.31 &[0.1, 0.9] \\ 
		$A_\mathrm{s}10^{9}$ & 2.08 & [$0.5$, $5.0$]  \\ 
		$n_{\rm s}$ & 0.96 & [0.87, 1.07]  \\
		$w$ &  -1.0  &[-2, -0.33]   \\
		$\Omega_{\rm b}$ & 0.049 &[0.03, 0.07]  \\
		$h_0$  & 0.67  &[0.55, 0.91]   \\
		$\Sigma\, m_{\nu}$ [eV]  & 0.06 & [0.0, 0.6]
		\\
        \midrule
	
		\multicolumn{3}{c}{\textbf{Linear galaxy bias  }}	 \\ [+0.1cm]
		$b^{i}$  & $1.35, 1.51, 1.65, 2.21, 2.44$ & [0.8,4.0]\\
        \midrule

		\multicolumn{3}{c}{\textbf{Lens
				magnification } } \\ [+0.1cm]
		$\alpha^{1} $ & 0.94 & ($0.94, 0.010$) \\
        $\alpha^{2} $ & 1.62 & ($1.62, 0.010$) \\
        $\alpha^{3} $ & 2.19 & ($2.19, 0.022$) \\
        $\alpha^{4} $ & 2.54 & ($2.54, 0.036$) \\
        $\alpha^{5} $ & 2.49 & ($2.49, 0.12$) \\ \midrule

		\multicolumn{3}{c}{{\bf
				Intrinsic alignments}} \\  [+0.1cm]    
        $A_{\rm IA}$   & 0.4 &  [$-5,5$ ]\\
		$\eta_{\rm IA}$   & 2.2  & [$-5,5$ ]\\\midrule

        \multicolumn{3}{c}{{\bf
				Baryonic feedback}} \\  [+0.1cm] 
        $\log_{10}(T_{\rm AGN}/K)$  &  7.8  & [$7.3, 8.3$]\\
        \bottomrule
	\end{tabular}
\end{table}

We infer the cosmological parameters from the measured data vectors using Bayesian statistics, estimating the posterior probability distribution for the parameters $\mathbf p$ of the theoretical model $M$, given the data $\hat{\mathbf D}$, as
\begin{equation}
    P(\mathbf p|\hat{\mathbf D}, M)\propto \mathcal{L}(\hat{\mathbf D}|\mathbf p, M) \, \Pi\left(\boldsymbol{\mathrm{p}}|M\right),
    \label{eq: bayes}
\end{equation}
where $\Pi\left(\boldsymbol{\mathrm{p}}|M\right)$ is the prior probability distribution for the parameters, and $\mathcal{L}(\hat{\mathbf D}|\mathbf p, M) $ is the likelihood of the data given a set of values for the parameters. We assume a Gaussian likelihood,
\begin{equation}
    \hspace{-0.5em} \ln \mathcal{L}(\hat{\mathbf D}|\mathbf p) = B -\dfrac{1}{2}\left(\hat{\mathbf D}-\mathbf T(\mathbf p)\right)^{\mathrm{T}} \mathbf{C}^{-1}\left(\hat{\mathbf D}-\mathbf T(\mathbf p)\right),
    \label{eq: likelihood}
\end{equation}
with a normalisation constant $B$, $T(\mathbf p)$ being the vector of theoretical predictions for each correlation function included in the data vector $\hat{\mathbf D}$, calculated assuming a given set of parameters $\mathbf p$, and $\mathbf{C}$ being the covariance matrix of the data, described in Sec.~\ref{subsec: cov}. 
The proportionality constant of Eq.~\eqref{eq: bayes} is given by the inverse of the Bayesian evidence:
\begin{equation}
    P(\hat{\mathbf D}|M) = \int \mathrm{d}\mathbf{p} \, \mathcal{L}(\hat{\mathbf D}|\mathbf p) \, \Pi\left(\boldsymbol{\mathrm{p}}|M\right).
\end{equation}
For some cases, we use the ratio of the Bayesian evidences to assess if the data prefer one model over the other (e.g. TATT over NLA). This approach has the advantage of penalising models with extra freedom in the parameter space. We use the Jeffreys scale \citep{evidence_ratio} to interpret the evidence ratio results.

Table~\ref{tab:params} lists the set of cosmological and astrophysical parameters, and their priors, considered in our analysis. We assume two main cosmological models, flat \lcdm\, and flat $w$CDM with free neutrino mass\footnote{We consider a variation with fixed neutrino mass in Sec.~\ref{subsec:robustness}.}. Regarding cosmological parameters, we sample the total matter energy density $\Om$, the amplitude of primordial scalar density fluctuations $A_{\rm s}$, the spectral index $n_{\rm s}$ of the power spectrum, the energy density of baryons $\Omega_{\rm b}$, the Hubble parameter $h$, and the sum of neutrino masses $\Sigma\, m_{\nu}$. In $w$CDM, this list is extended to sample $w$, for the equation of state of dark energy, which is assumed to be constant with time. In \lcdm\, this parameter is fixed to $-1$, the cosmological constant. In all cases, flatness is imposed by setting the dark energy density to $\Omega_{\Lambda}=1-\Om$. We follow \cite{2022PhRvD.105b3520A} for the choice of prior ranges for these parameters.  Although we sample the parameter $A_{\rm s}$, instead we report constraints on the rms amplitude of mass fluctuations on the $8\,h^{-1}\, \rm Mpc$  scale in linear theory, $\sigma_8$, or the related parameter $S_8\equiv \sigma_8(\Om/0.3)^{0.5}$, which is typically constrained best by weak lensing analyses.

In addition to the cosmological parameters, we marginalise over several parameters related to the galaxy bias, lens magnification, intrinsic alignments, and baryonic feedback effects. As described in Sec.~\ref{subsec:model}, we assume a linear galaxy bias model with one free amplitude parameter per redshift bin. For lens magnification, we also consider one amplitude parameter per redshift bin and assume Gaussian priors with mean and standard deviation from Table 3 of \cite{2025arXiv250621677H}, where these coefficients were measured for our BGS and LRG lens redshift bins. The NLA modelling of intrinsic alignments consists of two free parameters, for which we assume wide flat priors following previous analyses \citep{2023OJAp....6E..36D,2022PhRvD.105b3520A}. In Sec.~\ref{subsec:robustness} we also consider the TATT IA model. In that case, we have five free nuisance parameters, for which we assume the same priors as NLA for $A_1, \eta_1, A_2,$ and $\eta_2$, and the prior $[0, 2]$ for $b_{\rm TA}$ (following \citealt{2022PhRvD.105b3520A}).   
Last, we model the impact of baryonic feedback effects, such as baryon cooling, star formation, and active galactic nuclei (AGN) feedback, on the non-linear matter power spectrum through the free parameter $T_{\rm AGN}$ of \textsc{HMcode2020}, which has been calibrated with the BAHAMAS hydrodynamical simulations \citep{BAHAMAS2017,BAHAMAS2020}. $T_{\rm AGN}$ maps to the BAHAMAS-defined heating temperature of the AGN, modulating the strength of the baryon feedback suppression of the matter power spectrum. We adopt a top-hat prior on $\log_{10}(T_{\rm AGN}/K)$ of $[7.3, 8.3]$, where the lower limit follows \cite{2023OJAp....6E..36D}, and the upper limit is increased following \cite{DESY3blueshear-2024} to account for the recent evidence of stronger baryonic-feedback suppression \citep[see, e.g.,][]{Hadzhiyska2023, Bigwood2024}. 

 \begin{table}

    \caption{Priors for the nuisance parameters related to the calibration of the weak-lensing data from DES-Y3, KiDS-1000, and HSC-Y3. Square brackets denote a flat prior, while parentheses denote a Gaussian prior of the form $\mathcal{N}(\mu,\sigma)$. The Gaussian priors are, in general, uncorrelated between tomographic bins, except for the KiDS-1000 parameters related to shear $m^j$ and photometric redshift uncertainties $\Delta z^j$. In that case, the priors are correlated and modelled through a five-dimensional multivariate Gaussian prior with mean $\vek{\mu}$ and covariance $\vek{C}$. We list the diagonal of the covariance as $\sigma^j = \sqrt{C^{jj}}$, and denote these correlated priors as $(\mu^j, \sigma^j)_{\rm c}$.} 
    \label{tab: nuisance params}
	\centering	
	\begin{tabular}{cccc}
		\toprule
		Parameter & DES-Y3 & KiDS-1000 & HSC-Y3\\
        \midrule
		\multicolumn{4}{c}{\textbf{Photometric redshifts}} \\ [+0.1cm] 
		$\Delta z^1$ &  ($0.0, 0.018$) & $(0.000, 0.011)_{\rm c}$ & ($0.0, 0.024$)  \\ 
        $\Delta z^2$ &  ($0.0, 0.015$) & $(0.002, 0.011)_{\rm c}$ & ($0.0, 0.022$)  \\ 
        $\Delta z^3$ &  ($0.0, 0.011$) & $(0.013, 0.012)_{\rm c}$ & [$-1.0, 1.0$]  \\ 
        $\Delta z^4$ &  ($0.0, 0.017$) & $(0.011, 0.009)_{\rm c}$ & [$-1.0, 1.0$]  \\ 
        $\Delta z^5$ &  -              & $(-0.006, 0.010)_{\rm c}$ & -  \\ 
        \midrule

        \multicolumn{4}{c}{\textbf{Shear calibration}} \\ [+0.1cm] 
		$m^1$ &  ($-0.0063, 0.0091$) & $(-0.009, 0.019)_{\rm c}$ & ($0.00, 0.01$)  \\ 
        $m^2$ &  ($-0.0198, 0.0078$) & $(-0.011, 0.020)_{\rm c}$ & ($0.00, 0.01$)  \\ 
        $m^3$ &  ($-0.0241, 0.0076$) & $(-0.015, 0.017)_{\rm c}$ & ($0.00, 0.01$)  \\ 
        $m^4$ &  ($-0.0369, 0.0076$) & $(0.002, 0.012)_{\rm c}$  & ($0.00, 0.01$)  \\ 
        $m^5$ &  -                   & $(0.007, 0.010)_{\rm c}$  & -  \\
        $\delta_{\rm c}\times 10^4$ & -         & ($0.0, 2.3$)   &  - \\
        \midrule

        \multicolumn{4}{c}{\textbf{PSF systematics}} \\ [+0.1cm] 
		$\alpha_{\rm PSF}^{(2)}$ &  - & - & ($0.0, 1.0$)  \\ [+0.1cm]
        $\beta_{\rm PSF}^{(2)}$ &  - & -  & ($0.0, 1.0$)  \\ [+0.1cm]
        $\alpha_{\rm PSF}^{(4)}$ &  - & - & ($0.0, 1.0$)  \\ [+0.1cm]
        $\beta_{\rm PSF}^{(4)}$ &  - & -  & ($0.0, 1.0$)  \\ [+0.1cm]
        \bottomrule
	\end{tabular}
\end{table}

Table~\ref{tab: nuisance params} lists the data-calibration nuisance parameters and their priors as provided by each weak lensing survey: DES-Y3 (\citealt*{2022PhRvD.105b3514A,2022PhRvD.105b3515S}), KiDS-1000 \citep{2021A&A...645A.104A}, and HSC-Y3 \citep{2023PhRvD.108l3518L}. With these, we marginalise over the uncertainties related to photometric redshifts, shear calibration and, in the case of HSC, systematics associated with the modelling of the point-spread function (PSF). 

\paragraph{Photometric redshifts} We parametrise photometric redshift uncertainties via an additive shift to the mean redshift of each source tomographic bin $j$, given by $\Delta z^j$ such that,
\begin{equation}
    n^j(z) \rightarrow n^j(z-\Delta z^j).
    \label{eq: photoz bias}
\end{equation}
While DES-Y3 and HSC-Y3 assume these parameters to be independent, KiDS-1000 takes into account the correlated uncertainty between bins via a multivariate Gaussian prior. See Sec.~3.3 of \cite{2021A&A...646A.129J} for details.  

\paragraph{Shear calibration} The uncertainty associated with the shear estimation is typically parametrised by the shear calibration bias parameters $m^j$, which relate the measured ellipticity of  galaxies to the true shear. In terms of two-point correlation functions, we have
\begin{equation}
    \xi_{\pm}^{ij}(\theta) \rightarrow (1+m^i) (1+m^j) \, \xi_{\pm}^{ij}(\theta).
\end{equation}
We assume Gaussian priors for these parameters, as validated and provided by each weak lensing survey. In the case of KiDS-1000, these parameters are assumed to be 100\% correlated between bins. The correlated uncertainty is then included in the analytical covariance for the cosmic shear data vector (see Eq.~(37) in \citealt{2021A&A...646A.129J}). 
Additionally, when using $\xi_{+}$, KiDS-1000 considers an additive shear bias term, such that $\xi_{+} \rightarrow \xi_{+} + \delta_c^2$.

\paragraph{PSF systematics} HSC-Y3 additionally marginalises over four PSF-related nuisance parameters. $\alpha_{\rm PSF}^{(2)}$ and $\alpha_{\rm PSF}^{(4)}$ correspond to the PSF leakage bias by the 2nd and 4th moments of the PSF shapes, while $\beta_{\rm PSF}^{(2)}$ and $\beta_{\rm PSF}^{(4)}$ are related to the PSF modelling error in the 2nd and 4th-order moments. See Eqs.~(24-28) in \cite{2023PhRvD.108l3518L} for the full expression of these terms. Given that \cite{Zhang2023} found the bias on $\xi_{-}$ due to PSF systematics to be negligible, this uncertainty is only accounted for $\xi_{+}$. Following \cite{2023PhRvD.108l3518L}, we take into account the correlation between these four parameters in the likelihood inference.

We use the \textsc{CosmoSIS} framework for the modelling and likelihood evaluation, and its implementation of \textsc{Nautilus}\footnote{\url{https://nautilus-sampler.readthedocs.io/}} \citep{nautilus} to sample the posterior probability distribution of the parameters and estimate the Bayesian evidence. The configuration we use for \textsc{Nautilus} consists of $n_{\rm live} = 10000$ for the number of live points, $n_{\rm networks} = 16$ for the number of neural networks used in the estimator, and the discard-exploration mode (so-called ``nautilus-r'' in \citealt{nautilus}). We then analyse these samples with the \textsc{GetDist}\footnote{\url{https://getdist.readthedocs.io/}} package \citep{getdist-python}  to visualise the one- and two-dimensional posterior probability distributions of the parameters. 

When reporting parameter constraints in Sec.~\ref{sec:results}, we provide the mean of the 1D marginalised posterior distribution, along with a credible interval that encompasses 68\% of the marginal posterior density around the mean of the distribution. We use the \textsc{CosmoSIS} \texttt{postprocess} function to obtain these statistics. We also provide, in parentheses, the \textit{Maximum a Posteriori} (MAP) values for each constraint. To estimate these with high accuracy, we iteratively run the \textsc{MaxLike} sampler from \textsc{CosmoSIS} five times, starting with the maximum posterior value of the \textsc{Nautilus chain}. We use the \texttt{Powell} method with a tolerance of 0.05 and a maximum of 1000 iterations.

\subsection{Quantifying consistency and goodness of fit}
\label{subsec:gof}

We use the weighted least squares ($\chi^2$) statistic evaluated at the best-fit parameters, $\chi^2_{\min}$, to assess the goodness of fit of the model to the data points. We can then quantify the goodness of fit with $p(\chi^2 > \chi^2_{\min} |\, \nu_{\rm eff})$, which is the probability that a $\chi^2$ exceeds $\chi^2_{\min}$, assuming that it follows a $\chi^2$ distribution with an effective  number of degrees of freedom $\nu_{\rm eff}= N_{\rm D} - N_{\rm p, eff}$. Here,  $N_{\rm D}$ is the total number of data points, and $N_{\rm p, eff}$ is the effective number of free parameters, which is typically smaller than the total number of free parameters due to the use of informative priors and existing correlations between parameters. 

To estimate the effective number of free parameters, we follow Eq.~(29) of \cite{RaveriHu2019},
\begin{equation}
    N_{\rm p, eff} \equiv N_{\rm p} - \mathrm{tr}[\mathbf{C}_{\Pi}^{-1} \mathbf{C}_{P}],
    \label{eq: neff}
\end{equation}
where $N_{\rm p}$ is the total number of free parameters that are sampled, $\mathbf{C}_{\Pi}$ is the covariance of the prior probability distribution, and $\mathbf{C}_{P}$ is the covariance of the posterior distribution. We use \textsc{CosmoSIS} to sample the prior distribution and the \textsc{tensiometer}\footnote{\url{https://tensiometer.readthedocs.io}} package to estimate $N_{\rm p, eff}$ from Eq.~\eqref{eq: neff}.

To assess the consistency between two different analyses and, given that the posterior distributions might be non-Gaussian, we use the non-Gaussian estimates of parameter shifts provided by the \textsc{tensiometer} package. In particular, we use the kernel density estimator algorithm from \cite{Raveri2021} to estimate the probability distribution of parameter shifts $P(\mathbf{\Delta} \mathbf{\theta})$ and quantify the posterior mass above the iso-contour of "\emph{no shift}" ($\mathbf{\Delta \theta} = 0$), 
\begin{equation}
    \Delta = \int_{P(\mathbf{\Delta} \mathbf{\theta})>P(\mathbf{0})} \mathrm{d}\mathbf{\Delta \theta} \, P(\mathbf{\Delta} \mathbf{\theta}).
\end{equation}
We then convert this probability into an effective number of standard deviations $\sigma$ through $n_{\sigma,\rm eff}(P)=\sqrt{2} \, \rm Erf^{-1}(P)$, following \cite{RaveriHu2019}.  We have compared these estimates with those from the Gaussian estimator $Q_{\rm DM}$, defined in \cite{RaveriHu2019}, finding practically the same statistical significance for the main cosmological parameters of interest, $\Om$ and $S_8$. 

\subsection{Scale cuts}
\label{subsec: scale cuts}

In an accompanying paper \citep{Nimas2025}, we validate our modelling and scale cut choices for the fits to the data (for either cosmic shear, the combination of projected galaxy clustering and galaxy-galaxy lensing, or $2\times2$-pt, and the full $3\times2$-pt analysis).  Starting from the fiducial scale cuts for cosmic shear from each survey, we follow the methodology of \cite{Krause2021} to ensure that our modelling is robust (in the scales included) against uncertainties in the modelling of the non-linear matter power spectrum and non-linear galaxy bias.  We therefore remove the smallest scales from our analysis, to mitigate the sensitivity to these modelling uncertainties.

The methodology consists of comparing the cosmology contours between synthetic data vectors generated with either our \emph{``baseline''} model or a \emph{``contaminated''} model that includes an additional or alternative effect not included in our baseline choice. To ensure robustness against a given ``contamination'' or modelling choice, we set a threshold of $\leq0.3\sigma$ in the $S_8-\Om$ 2D plane (and $w-\Om$ for $w$CDM) for the difference between the baseline and contaminated posterior distributions.  The ``contaminating'' effects we consider are non-linear galaxy bias, baryonic feedback effects imprinted in the matter power spectrum, and an alternative non-linear matter power spectrum prediction. To generate these synthetic data vectors, we rely on the hybrid effective field theory (HEFT) \citep{Modi2020} implementation of the \textsc{Aemulus} $\nu$ emulator from \cite{DeRose2023}, which has been used in a previous $2\times2$-pt analysis of DESI-DR1 and DES-Y3 data \citep{Chen2024}. 

After testing against these modelling effects for all probe and survey combinations, we find that the fiducial scale cuts for DES-Y3 (\citealt*{2022PhRvD.105b3514A,2022PhRvD.105b3515S}) and for HSC-Y3 \citep{2023PhRvD.108l3518L} cosmic shear are still valid for our analysis. For KiDS-1000 cosmic shear, however, we find that the fiducial scales from \cite{2021A&A...645A.104A} marginally do not satisfy our threshold criteria. In this case, we follow the methodology from \cite{Krause2021} for cosmic shear to decide which data points to remove first from the analysis, based on the $\chi^2$ difference between baseline and contaminated predictions for each pair-bin combination $\xi_{\pm}^{ij}$. We found that we could meet our $0.3\sigma$ threshold criteria by changing the $\xi_{-}$ scale cuts from $4'$ to $6.06'$, which just removes one additional data point in some bin combinations with respect to \cite{2021A&A...645A.104A}. The fiducial scale cuts for $\xi_{+}$ ($0.5'$) remain unchanged.     

The scale cuts for galaxy-galaxy lensing and projected galaxy clustering are defined in terms of the projected separation $r_{\rm p}$, which we convert to angular scales for $\gamma_{\rm t}$ using $\theta_{\min}^i=r_{\rm p,\min}/\chi(z_{\rm lens}^i)$, with $\chi(z_{\rm lens}^i)$ being the value of the comoving distance at the effective redshift of lens bin $i$. We find that the scale cuts $(r_{\rm p,ggl},\, r_{\rm p,clus}) = (6,\, 8 )\,h^{-1}\,\mathrm{Mpc}$ for galaxy-galaxy lensing and galaxy clustering, respectively, satisfy our validation criteria for all combinations. We note that these scales correspond to those used in the DES-Y3 analysis for the linear galaxy bias model \citep{Krause2021,desy3_2x2pt,2022PhRvD.105b3520A}.

\subsection{Covariance}
\label{subsec: cov}

We now summarise the covariance model we adopt for our set of $3 \times 2$-pt correlation functions $\{ \xi_\pm(\theta), \gamma_t(\theta), w_p(r_p) \}$.  The foundation of our model is a Gaussian analytical covariance including sample variance, noise and mixed terms.  Our implementation of these terms for the auto- and cross-covariances of the different 2-pt correlations, between different source and lens samples, is fully presented in Appendices A-D of \cite{2025OJAp....8E..24B}; our computations follow frameworks previously presented by \cite{2017MNRAS.470.2100K} and \cite{2021A&A...646A.129J}.  For the cosmic shear covariance we include super-sample variance \citep{2013PhRvD..87l3504T}, and we also apply a small-scale noise correction to all covariances based on the number of observed pairs within the survey geometry \citep{2018MNRAS.479.4998T}.  We corrected the galaxy-galaxy lensing covariance for footprint boundary effects using an ensemble of fast mocks sampled to the angular selection function of the lensing surveys, as discussed by \cite{2024MNRAS.533..589Y}.

We validated our analytical covariance pipeline through comparison with three existing codes: \textsc{cosmocov} \citep{2017MNRAS.470.2100K, 2020MNRAS.497.2699F} used by DES, the code used by KiDS-1000 \citep{2021A&A...646A.129J}, and the \textsc{onecovariance} framework \citep{2025A&A...699A.124R} employed by the KiDS-Legacy analysis \citep[these first two validations are discussed in Appendix E of][]{2025OJAp....8E..24B}.  We specifically used \textsc{onecovariance} to validate that the non-Gaussian contribution to the cosmic shear and galaxy-galaxy lensing covariance was negligible for our configurations (i.e., had no significant effect on $\chi^2$ values).  We also ran test cosmology fits in which we replaced our fiducial cosmic shear covariance with the equivalent versions generated by the DES-Y3, KiDS-1000 and HSC-Y3 collaborations \citep{2021MNRAS.508.3125F, 2021A&A...646A.129J, 2023PhRvD.108l3518L}, validating that our cosmological results in these cases were not significantly changed.
For the final cosmology results from Sec.~\ref{sec:results} we use a jack-knife covariance for $w_{\rm p}$, estimated in \cite{2025arXiv250621677H}, instead of the analytical prediction. This change does not impact the final cosmological constraints, but it significantly improves the goodness-of-fit statistics, as discussed in Sec.~\ref{sec:results}. The reason for this is that the analytical covariance estimate for $w_{\rm p}$ does not currently include the footprint boundary effects.

Our shear and clustering datasets cover different sky areas, which only partially overlap.  We corrected our covariance model for this effect by assuming that correlation measurements in non-overlapping regions are uncorrelated \citep[which was found to be a safe assumption by][]{2023OJAp....6E..36D}.  If the sky area coverage of two datasets is $\Omega_1$ and $\Omega_2$, with overlap area $\Omega_O$, then the covariance of the correlation functions of these datasets should be scaled by $\Omega_O^2/\Omega_1 \Omega_2$ in comparison with fully-overlapping measurements \citep{2025OJAp....8E..24B}.  The full set of areas assumed in our covariance model is reproduced in Table \ref{tab:areas}.

\begin{table}
    \caption{The areas of the DESI-DR1 and weak lensing surveys used in our covariance model, and their intersections, in square degrees.  The row and column labelled ``Full'' provide the area of the individual components, and the remainder of the table quotes the areas of intersection.  The DESI sample is divided into BGS, LRG, and the common area (BGS+LRG).  The ``Full'' areas for the lensing surveys are quoted from the survey collaborations, and the remaining areas are evaluated on an $n_{\rm side} = 1024$ {\sc healpix} grid.} 
    \label{tab:areas}
	\centering	
	\begin{tabular}{lcccc}
		\toprule
		WL survey & Full & DESI-BGS & DESI-LRG & DESI-BGS+LRG \\
        \midrule
        Full & -- & 9146.5 & 8607.6 & 6285.5 \\
        DES-Y3 & 4143.0 & 716.8 & 851.2 & 587.5 \\
        KiDS-1000 & 773.3 & 448.2 & 446.8 & 446.3 \\
        HSC-Y1 & 136.9 & 142.1 & 151.8 & 139.0 \\
        HSC-Y3 & 417.0 & 440.6 & 440.5 & 413.3 \\
        \bottomrule
	\end{tabular}
\end{table}

\subsection{Blinding}
\label{subsec: blinding}

We implement a blinding procedure to safeguard our analysis against cognitive biases such as confirmation bias. To this end, the goal of our blinding procedure is to randomly shift cosmological parameter constraints in our analysis. This allows us to perform a variety of robustness tests without potentially being influenced by expectations about what our analysis should reveal in terms of cosmology. Only after passing the robustness test do we run the analysis on unblinded data and obtain our actual cosmological constraints.

Our blinding procedure is based on altering the data vectors, i.e., the $3 \times 2$-pt correlation functions $\xi_\mathrm{D} = \{ \xi_\pm, \gamma_{\rm t}, w_{\rm p} \}$. Given a random set of shifts in the cosmological parameters, $\Delta \mathcal{C}$, the blinding method produces a new blinded data vector $\tilde \xi_\mathrm{D} = \xi_\mathrm{D} + \Delta \xi$ whose posterior is roughly shifted by $\Delta \mathcal{C}$ with respect to that of $\xi_\mathrm{D}$. At the same time, the blinded data vector $\tilde \xi_\mathrm{D}$ should be equally sensitive to robustness tests, e.g., should result in a similar $\chi^2$ with respect to the best-fit model $\tilde \xi_\mathrm{T}$. \cite{Muir2020_MNRAS_494_4454} introduced the following blinding function,
\begin{equation}
    \Delta \xi = \xi_\mathrm{T}(\mathcal{C}_\mathrm{bli}, \mathcal{N}_\mathrm{fid}) - \xi_\mathrm{T}(\mathcal{C}_\mathrm{fid}, \mathcal{N}_\mathrm{fid}) \, .
\end{equation}
In the above equation, $\mathcal{C}_\mathrm{fid}$ represents a choice of fiducial cosmological parameters, $\mathcal{C}_\mathrm{bli} = \mathcal{C}_\mathrm{fid} + \Delta \mathcal{C}$ and $\mathcal{N}_\mathrm{fid}$ are fiducial nuisance parameters such as galaxy bias parameters. \cite{Muir2020_MNRAS_494_4454} showed that, when applied to the DES-Y3 analysis, this simple procedure shifts the cosmological posterior of $\tilde \xi_\mathrm{D}$ by roughly $\Delta \mathcal{C}$ while leaving the goodness of fit largely unchanged. In this work, we implement a slightly modified version of that blinding approach:
\begin{equation}
    \Delta \xi = \xi_\mathrm{T}(\mathcal{C}_\mathrm{bli}, \mathcal{N} (\mathcal{C}_\mathrm{bli})) - \xi_\mathrm{T}(\mathcal{C}_\mathrm{fid}, \mathcal{N}(\mathcal{C}_\mathrm{fid})) \, .
\end{equation}
In the above equation, compared to \cite{Muir2020_MNRAS_494_4454}, we do not choose a fiducial set of nuisance parameters but instead optimise them with respect to the cosmological parameters. In particular, the nuisance parameters $\mathcal{N}(\mathcal{C})$ are chosen so that they minimise the squared Mahalanobis distance, i.e., the $\chi^2$, between the data, $\xi_\mathrm{D}$, and the model, $\xi_\mathrm{T}$ for a fixed cosmology $\mathcal{C}$. The motivation for this slightly different procedure is that it minimises an upper bound on the shift $\Delta \xi$, i.e.,

\begin{equation}
    \begin{split}
        \| \Delta \xi \|^2 \leq &2 \| \xi_\mathrm{T}(\mathcal{C}_\mathrm{bli}, \mathcal{N} (\mathcal{C}_\mathrm{bli})) - \xi_\mathrm{D} \|^2 + \\
        &2 \| \xi_\mathrm{T}(\mathcal{C}_\mathrm{fid}, \mathcal{N} (\mathcal{C}_\mathrm{fid})) - \xi_\mathrm{D} \|^2 \, .
    \end{split}
\end{equation}

In other words, the data vector shift $\Delta \xi$ is guaranteed to be small since both theory data vectors are similar to the observed data vector. In practice, in this work, we only blind the cosmological parameters $\Omega_\mathrm{m}$ and $\sigma_8$ and only optimise the galaxy bias parameters when it comes to nuisance parameters. All other cosmological and nuisance parameters are left at their fiducial choice. In particular, we choose $\Omega_\mathrm{m, fid} = 0.3166$ and $\sigma_{8, \mathrm{fid}} = 0.8120$ and draw $\Delta \Omega_\mathrm{m}$ and $\Delta \sigma_8$ from uniform distributions with ranges $[-0.05, +0.05]$ and $[-0.12, +0.12]$, respectively. We tested that this procedure works reliably on mock data, i.e., produces shifts in line with the expected shift $\Delta \mathcal{C}$ while not leading to significant shifts in the best-fit $\chi^2$.

\section{Cosmological results}
\label{sec:results}

\begin{figure}
    \centering
    \includegraphics[width=\linewidth]{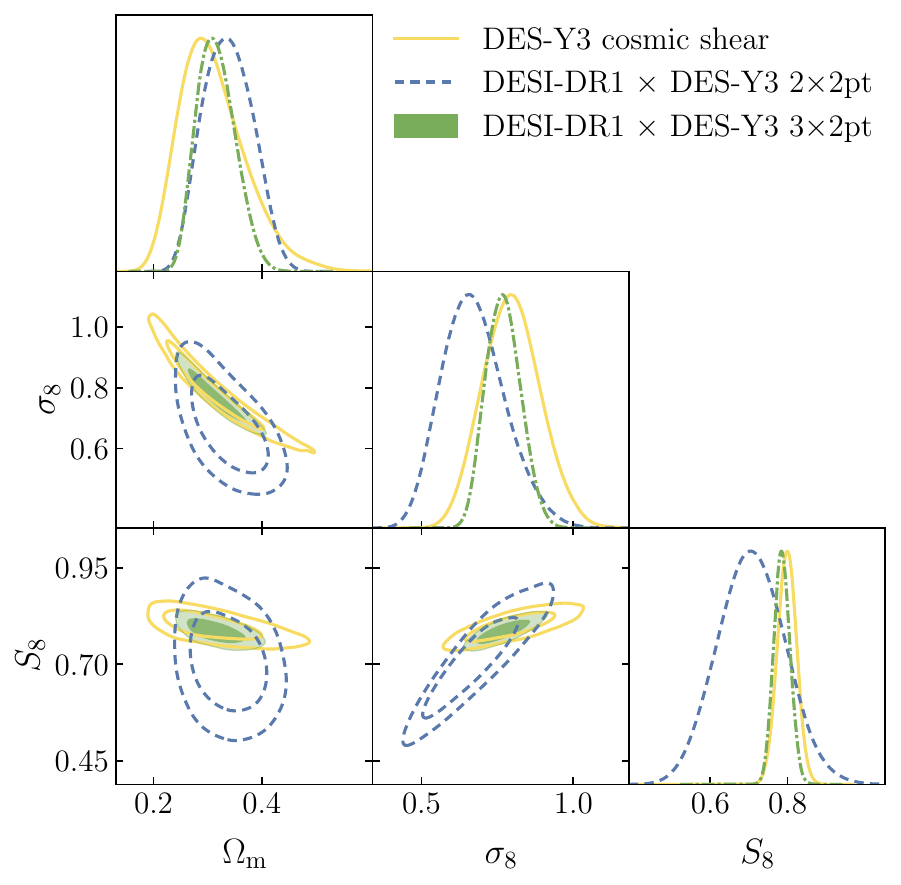}
    \caption{Marginalised constraints on $\Om$, $\sigma_8$, and $S_8$ in \lcdm\, for DES-Y3 cosmic shear (solid yellow), the combination of galaxy-galaxy lensing and projected galaxy clustering from DESI-DR1 (dashed blue), and the $3\times2$-pt combination (solid green).}
    \label{fig:desy3_lcdm}
\end{figure}

\begin{figure}
    \centering
    \includegraphics[width=\linewidth]{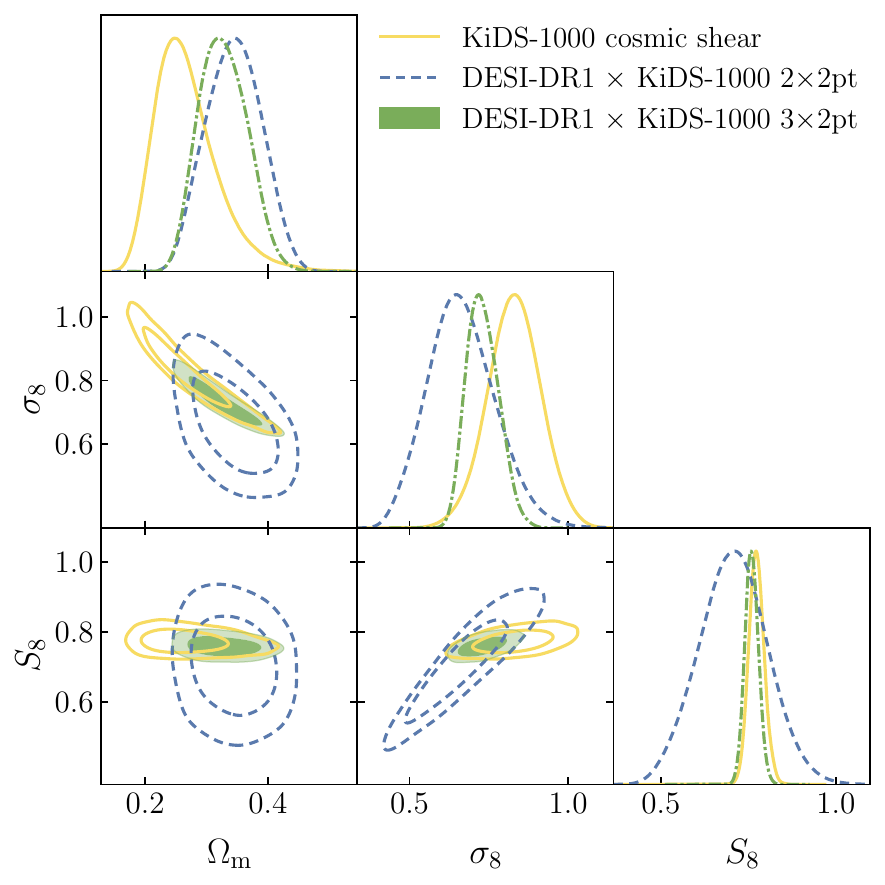}
    \caption{Marginalised constraints on $\Om$, $\sigma_8$, and $S_8$ in \lcdm\, for KiDS-1000 cosmic shear (solid yellow), the combination of galaxy-galaxy lensing and projected galaxy clustering from DESI-DR1 (dashed blue), and the $3\times2$-pt combination (solid green).}
    \label{fig:kids1000_lcdm}
\end{figure}

\begin{figure}
    \centering
    \includegraphics[width=\linewidth]{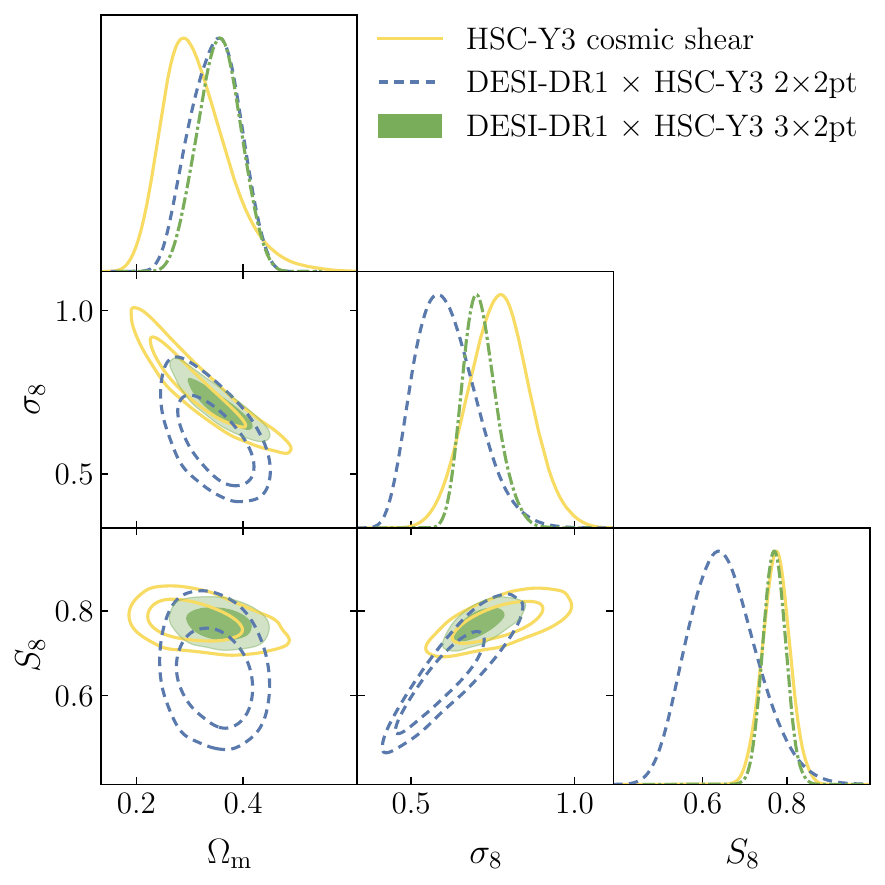}
    \caption{Marginalised constraints on $\Om$, $\sigma_8$, and $S_8$ in \lcdm\, for HSC-Y3 cosmic shear (solid yellow), the combination of galaxy-galaxy lensing and projected galaxy clustering from DESI-DR1 (dashed blue), and the $3\times2$-pt combination (solid green).}
    \label{fig:hscy3_lcdm}
\end{figure}

\begin{table*}
    \caption{Constraints on $S_8$, $\Om$, and $\sigma_8$ in \lcdm\, with 68\% credible intervals using the mean 1D marginal posterior distribution. For each weak lensing dataset, we report constraints on cosmic shear, the combination of DESI-DR1 projected galaxy clustering and galaxy-galaxy lensing ($2\times2$-pt), and the full $3\times2$-pt data vector. Along with the marginalised constraints, we provide goodness of fit statistics: the $\chi^2$ value at the best fit ($\chi^2_{\min}$), the total number of data points $N_{\rm D}$, the estimated effective number of free parameters $N_{\rm p, eff}$, the reduced $\chi^2_{\rm red}=\chi^2_{\min}/(N_{\rm D}-N_{\rm p, eff})$, and the goodness of fit probability $p$ (see Sec.~\ref{subsec:gof} for more details).  
    } 
    \label{tab:lcdm-constraints}
	\centering	
	\begin{tabular}{lcccccccc} 
		\toprule
		Analysis & $S_8$ & $\Om$ & $\sigma_8$ & $\chi^2_{\min}$ & $N_{\rm D}$ & $N_{\rm p, eff}$ & $\chi^2_{\rm red}$ & $p(\chi^2_{\rm red})$\\
        \midrule
		\multicolumn{8}{l}{\textbf{DES-Y3}} \\ [+0.1cm] 
        Cosmic shear & $0.799^{+0.025}_{-0.025}$ & $0.311^{+0.074}_{-0.043}$ &$0.797^{+0.104}_{-0.090}$ & 241.72  & 227  & 4.41 & 1.09 & 0.18   \\ [+0.1cm] 
        $2\times2$-pt & $0.707^{+0.090}_{-0.086}$ & $0.337^{+0.049}_{-0.046}$ &$0.673^{+0.115}_{-0.089}$& 131.55 & 137  & 8.86  & 1.03 & 0.40   \\ [+0.1cm]
        $3\times2$-pt & $0.786^{+0.022}_{-0.019}$ & $0.316^{+0.040}_{-0.031}$ &$0.770^{+0.062}_{-0.054}$ & 396.02  &  364  & 10.19 &  1.12 & 0.06   \\
        \midrule

        \multicolumn{8}{l}{\textbf{KiDS-1000}} \\ [+0.1cm] 
        Cosmic shear & $0.773^{+0.023}_{-0.020}$ & $0.268^{+0.060}_{-0.032}$ &$0.830^{+0.081}_{-0.082}$ & 237.22  & 221 & 3.90 & 1.09 & 0.17   \\ [+0.1cm] 
        $2\times2$-pt & $0.707^{+0.094}_{-0.098}$ & $0.345^{+0.046}_{-0.046}$ &$0.665^{+0.116}_{-0.096}$& 163.47 & 165  & 8.61 & 1.05 & 0.33   \\ [+0.1cm]
        $3\times2$-pt & $0.760^{+0.020}_{-0.018}$ & $0.327^{+0.043}_{-0.036}$ &$0.731^{+0.059}_{-0.045}$ & 414.84 & 386 & 9.80 & 1.10 & 0.08   \\[+0.1cm]
        \midrule

        \multicolumn{8}{l}{\textbf{HSC-Y3}} \\ [+0.1cm] 
        Cosmic shear & $0.775^{+0.031}_{-0.032}$ & $0.313^{+0.072}_{-0.044}$ &$0.770^{+0.094}_{-0.091}$ & 141.63 & 140  & 6.47 & 1.06 & 0.30  \\ [+0.1cm] 
        $2\times2$-pt & $0.649^{+0.091}_{-0.067}$ & $0.347^{+0.046}_{-0.048}$ &$0.609^{+0.110}_{-0.072}$& 99.71 & 137  & 11.34 & 0.79 & 0.55  \\ [+0.1cm]
        $3\times2$-pt & $0.771^{+0.026}_{-0.027}$ & $0.355^{+0.040}_{-0.040}$ &$0.712^{+0.060}_{-0.042}$ & 246.56 & 277  & 12.85 & 0.93 & 0.77  \\

        \bottomrule
	\end{tabular}
\end{table*}

In this section, we provide cosmological results for the \lcdm\, (Sec.~\ref{subsec:lcdm-results}) and $w$CDM models (Sec.~\ref{subsec:wcdm-results}) for each weak lensing survey and probe combination. In Sec.~\ref{subsec:robustness}, we then carry out several analysis variations to assess the robustness of our main results. 

We first analysed the blinded data vectors from Sec.~\ref{subsec: blinding} and ensured that: (1) the best-fit goodness of fit $\chi^2$ has $p>0.01$ for each of the results (including combinations such as $\xi_{\pm}+w_{\rm p}$ and $\xi_{\pm} + \gamma_{\rm t}$, to identify potential issues with individual observables), (2) the posteriors of nuisance parameters with Gaussian priors from Table~\ref{tab: nuisance params} are not hitting the edges of the priors, and (3) the cosmic shear and $2\times2$-pt constraints for each WL survey are internally consistent (see Sec.~\ref{subsec:gof} for details).

Our first blinded results provided low goodness-of-fit $p$ values for most cases. After comparison with the fiducial covariances from the cosmic shear analyses, we corrected the covariances for cosmic shear and galaxy-galaxy lensing by accounting for the exact number of pairs \citep{2018MNRAS.479.4998T} (see Sec.~\ref{subsec: cov}). After this correction, we still found low $p$ values when including $w_{\rm p}$ in the likelihood inference, which was resolved after replacing the $w_{\rm p}$ analytical covariance, which is the least accurate of the observables due to footprint effects, with the jack-knife covariance estimated in \cite{2025arXiv250621677H}. After these changes, we still found low goodness-of-fit $p$ values for some combinations using HSC Year 1 data. We report cosmological constraints with HSC-Y1 in Appendix~\ref{sec: hsc-y1} for completeness. We note that the changes to the covariance did not significantly affect the resulting fitted parameter values.

\subsection{\texorpdfstring{$\Lambda$CDM}{LCDM}}
\label{subsec:lcdm-results}

We have obtained cosmological constraints using projected galaxy clustering from DESI-DR1 and weak lensing data from DES-Y3, KiDS-1000, and HSC-Y3. We show the marginalised posterior distributions for the main constrained cosmological parameters, $\Om$, $S_8$, and $\sigma_8$, in Figs.~\ref{fig:desy3_lcdm}, \ref{fig:kids1000_lcdm}, and \ref{fig:hscy3_lcdm}. In Table~\ref{tab:lcdm-constraints}, we report the 68\% credible intervals around the mean 1D marginal posterior distribution for these parameters, along with the goodness of fit metrics for each case. Our best constrained parameter is $S_8$ when using the whole $3\times2$-pt data vector:
\begin{equation}
    \begin{gathered}
    \nonumber
    \mathrm{DESI\text{-}DR1\, \times\, DES\text{-}Y3:}\qquad S_8= 0.786^{+0.022}_{-0.019} \,\,(0.790), \\
    \mathrm{DESI\text{-}DR1\, \times\, KiDS\text{-}1000:}\quad S_8= 0.760^{+0.020}_{-0.018}\,\, (0.776), \\
    \mathrm{DESI\text{-}DR1\, \times\, HSC\text{-}Y3:}\qquad  S_8= 0.771^{+0.026}_{-0.027}\,\, (0.763),
    \end{gathered}
\end{equation}
corresponding to 2.6\% (DES-Y3), 2.5\% (KiDS-1000), and 3.4\% (HSC-Y3) precision measurements. 

We provide additional posterior distributions for the \threex\, analyses in Appendix~\ref{sec:posteriors}. Besides the main cosmological parameters, with DES-Y3 and KiDS-1000 data, we are able to constrain the amplitude of intrinsic alignments under our fiducial NLA model,
\begin{equation}
    \begin{gathered}
    \nonumber
    \mathrm{DESI\text{-}DR1\, \times\, DES\text{-}Y3:}\qquad A_{\rm IA}= 0.36^{+0.49}_{-0.27}\,\,(0.43), \\
    \mathrm{DESI\text{-}DR1\, \times\, KiDS\text{-}1000:}\quad A_{\rm IA}= 0.55^{+0.54}_{-0.28}\,\, (0.08),
    \end{gathered}
\end{equation}
finding values very consistent with the constraints obtained for these cosmic shear data sets in \cite{2023OJAp....6E..36D}. In the case of HSC-Y3, we find unconstrained posteriors for the intrinsic alignment parameters that hit the edges of the priors, probably due to the fact that the data have to self-calibrate the photometric redshift parameters in source bins 3 and 4, for which we assume wide flat priors (see Table~\ref{tab: nuisance params}). For these $\Delta_z$ parameters, we obtain:
\begin{equation}
    \begin{gathered}
    \nonumber
    \mathrm{DESI\text{-}DR1\, \times\, HSC\text{-}Y3:}\quad \Delta_z^3= 0.068^{+0.041}_{-0.039}\,\, (0.071), \\
    \mathrm{DESI\text{-}DR1\, \times\, HSC\text{-}Y3:}\quad \Delta_z^4= 0.166^{+0.071}_{-0.069}\,\, (0.179).
    \label{eq: hsc deltaz constraints}
    \end{gathered}
\end{equation}
These constraints are consistent within 1$\sigma$ with the results reported by the HSC Y3 fiducial analyses \citep{2023PhRvD.108l3518L,2023PhRvD.108l3519D}\footnote{Note that the sign convention for $\Delta_z$ we assume in Eq.~\eqref{eq: photoz bias} is different to what is assumed in the official HSC Y3 analysis. A positive value here corresponds to a negative value in \cite{2023PhRvD.108l3518L}.}, but tighter by 27\% and 20\%, respectively, thanks to the use of DESI-DR1 data as lenses in this \threex\, analysis. 
Since then, other works in the literature have obtained constraints on these parameters (\citealt{Rana2026}, \citealt{Zhang2025}, and \citealt{Choppin2025,Choppin2025b}), which are in general consistent with our constrains at the $1\sigma$ level, with the exception of the $\Delta_z^4$ constraint from \cite{Choppin2025}, which is consistent within $2\sigma$. We note that \cite{Choppin2025} has done an independent calibration of the HSC Y3 redshift distributions, with additional DESI data that is not considered in this work. Here, we just give freedom to these parameters to self-calibrate, obtaining posterior distributions that are correlated between redshift bins and with the intrinsic alignment parameters (see Fig.~\ref{fig: hscy3 IA delta z}). As a result, there could be shifts along the degeneracy directions for poorly constrained parameters in the analysis. In any case, all these constraints are well within the priors assumed in our analysis, so we do not expect a significant change in our results from the different $\Delta_z$ values in bins 3 and 4.

\subsection{\texorpdfstring{$w$CDM}{wCDM}}
\label{subsec:wcdm-results}

\begin{table*}
    \caption{Constraints on $S_8$, $\Om$, and $\sigma_8$ in $w$CDM\, with 68\% credible intervals using the mean 1D marginal posterior distribution. For each weak lensing dataset, we report constraints on the $3\times2$-pt combination with DESI-DR1 projected galaxy clustering. Along with the marginalised constraints, we provide goodness of fit statistics: the $\chi^2$ value at the best fit ($\chi^2_{\min}$), the total number of data points $N_{\rm D}$, the estimated effective number of free parameters $N_{\rm p, eff}$, the reduced $\chi^2_{\rm red}=\chi^2_{\min}/(N_{\rm D}-N_{\rm p, eff})$, and the goodness of fit probability $p$ (see Sec.~\ref{subsec:gof} for more details).} 
    \label{tab:wcdm-constraints}
	\centering	
	\begin{tabular}{lccccccccc} 
		\toprule
		WL survey & $S_8$ & $\Om$ & $\sigma_8$ & $w$ & $\chi^2_{\min}$ & $N_{\rm D}$ & $N_{\rm p, eff}$ & $\chi^2_{\rm red}$ & $p(\chi^2_{\rm red})$\\
        \midrule
        DES-Y3 & $0.777^{+0.045}_{-0.039}$ & $0.316^{+0.043}_{-0.032}$ &$0.760^{+0.064}_{-0.057}$ & $-1.12^{+0.32}_{-0.45}$  & 395.87  & 364  & 10.26 & 1.12 & 0.06   \\ [+0.1cm] 
        KiDS-1000 & $0.770^{+0.030}_{-0.035}$ & $0.336^{+0.043}_{-0.043}$ &$0.732^{+0.057}_{-0.045}$& $-0.88^{+0.18}_{-0.42}$  & 413.50 & 386  & 10.04  & 1.10 & 0.09   \\ [+0.1cm]
        HSC-Y3 & $0.753^{+0.036}_{-0.034}$ & $0.357^{+0.038}_{-0.044}$ &$0.694^{+0.063}_{-0.045}$ & $-1.27^{+0.33}_{-0.40}$   & 246.52  &  277  & 13.18 &  0.93 & 0.77   \\

        \bottomrule
	\end{tabular}
\end{table*}

\begin{figure}
    \centering
    \includegraphics[width=\linewidth]{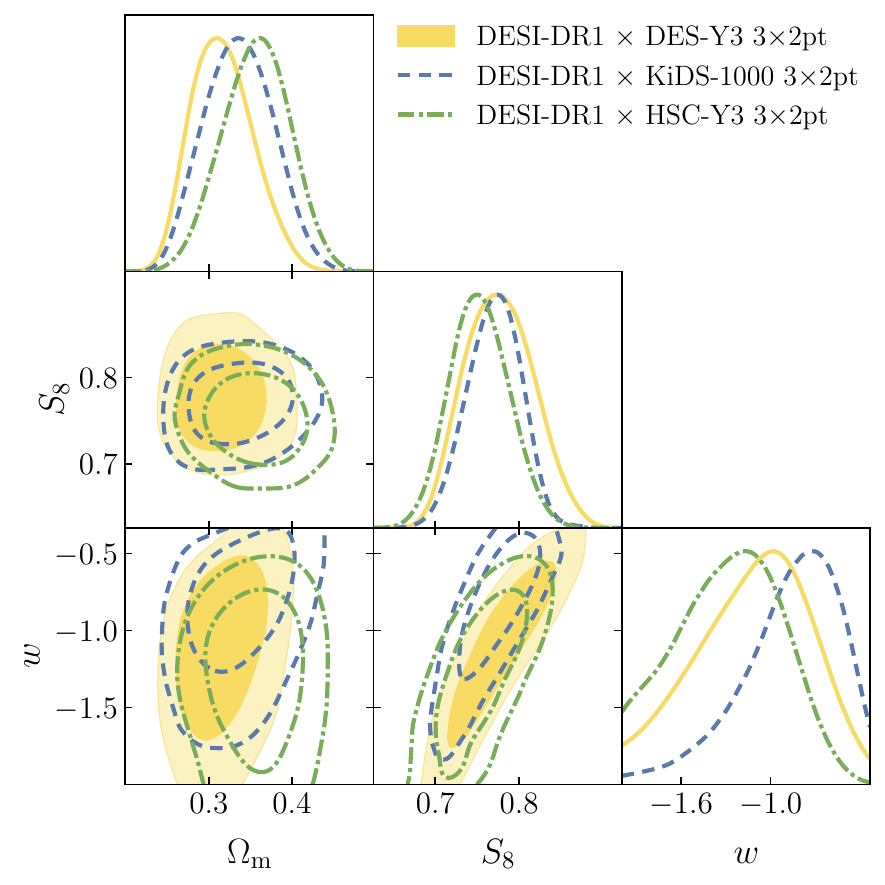}
    \caption{Marginalised constraints on $\Om$, $S_8$, and $w$ in $w$CDM from the  $3\times2$-pt combination of DESI-DR1 projected galaxy clustering and weak lensing data from DES-Y3 (solid yellow), KiDS-1000 (dashed blue), and HSC-Y3 (dash-dotted green).}
    \label{fig:wcdm}
\end{figure}

We now use our most constraining data combination, $3\times2$-pt, to constrain the equation of state of dark energy, $w$, within the $w$CDM model. Fig.~\ref{fig:wcdm} shows the marginalised posterior distributions of $w$, $\Om$, and $S_8$ for each survey combination. Similarly to the \lcdm\, case, we provide the 68\% credible intervals around the 1D mean marginal in Table~\ref{tab:wcdm-constraints}, including the constraints for $\sigma_8$ and the goodness-of-fit metrics. We find that the constraints are consistent between surveys and with the \lcdm\, expectation of $w=-1$,
\begin{equation}
    \begin{gathered}
    \nonumber
    \mathrm{DESI\text{-}DR1\, \times\, DES\text{-}Y3:}\qquad\, w= -1.12^{+0.32}_{-0.45}\,\, (-0.96), \\
    \mathrm{DESI\text{-}DR1\, \times\, KiDS\text{-}1000:}\quad w=-0.88^{+0.18}_{-0.42}\,\,(-0.87), \\
    \mathrm{DESI\text{-}DR1\, \times\, HSC\text{-}Y3:}\qquad \,  w= -1.27^{+0.33}_{-0.40}\,\,(-1.31).
    \end{gathered}
\end{equation}

\begin{table}
	\caption{Constraints on $S_8$ in $\Lambda$CDM\, assuming the TATT model for intrinsic alignments. We report the 68\% credible intervals using the mean 1D marginal posterior distribution with the MAP estimate between parentheses. For each weak lensing dataset, we provide constraints from the $3\times2$-pt combination with DESI-DR1 projected galaxy clustering. Along with the marginalised constraints, we provide the Bayesian evidence ratio of TATT to NLA ($R$) and goodness of fit statistics: the $\chi^2$ value at the best fit ($\chi^2_{\min}$), the estimated effective number of free parameters $N_{\rm p, eff}$, and the goodness of fit probability $p$ (see Sec.~\ref{subsec:gof} for more details).} 
	\label{tab:tatt-constraints}
	\centering	
	\begin{tabular}{lccccc} 
		\toprule
		WL survey & $S_8$ & $R$ [$\times10^{-2}$] & $\chi^2_{\min}$  & $N_{\rm p, eff}$  & $p(\chi^2_{\rm red})$\\
		\midrule
		\multirow{2}{*}{DES-Y3} & $0.780^{+0.023}_{-0.021}$ & \multirow{2}{*}{$7.2 \pm 0.1$} & \multirow{2}{*}{396.53}    & \multirow{2}{*}{11.27} & \multirow{2}{*}{0.05}   \\[+0.1cm]  
		& ($0.811$) &  & & &	\\
		\midrule
		\multirow{2}{*}{KiDS-1000} & $0.731^{+0.023}_{-0.025}$ & \multirow{2}{*}{$4901 \pm 59$}  & \multirow{2}{*}{399.20}   & \multirow{2}{*}{12.30}   & \multirow{2}{*}{0.17}   \\ [+0.1cm]
		& ($0.735$) &  & & &	\\
		\midrule
		\multirow{2}{*}{HSC-Y3} & $0.767^{+0.028}_{-0.025}$   & \multirow{2}{*}{$16.5 \pm 0.2$}  & \multirow{2}{*}{246.24}   & \multirow{2}{*}{13.91} & \multirow{2}{*}{0.76}   \\
		& ($0.769$) &  & & &	\\		
		\bottomrule
	\end{tabular}
\end{table}

\begin{figure*}
    \centering
    \includegraphics[width=0.9\linewidth]{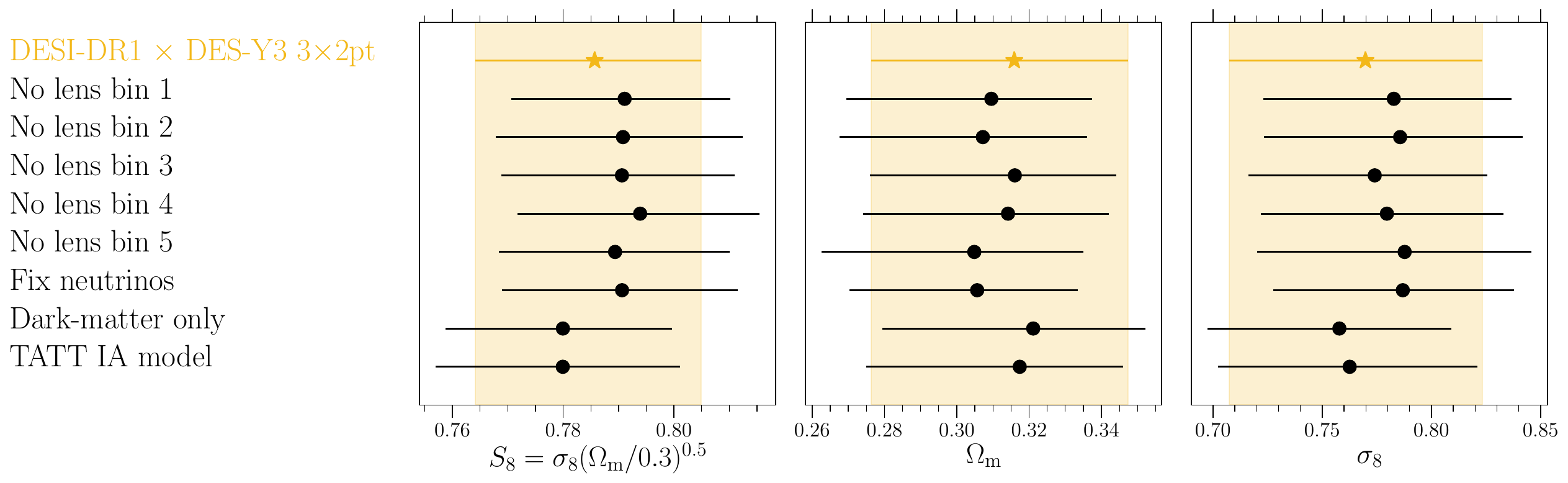}
    \includegraphics[width=0.9\linewidth]{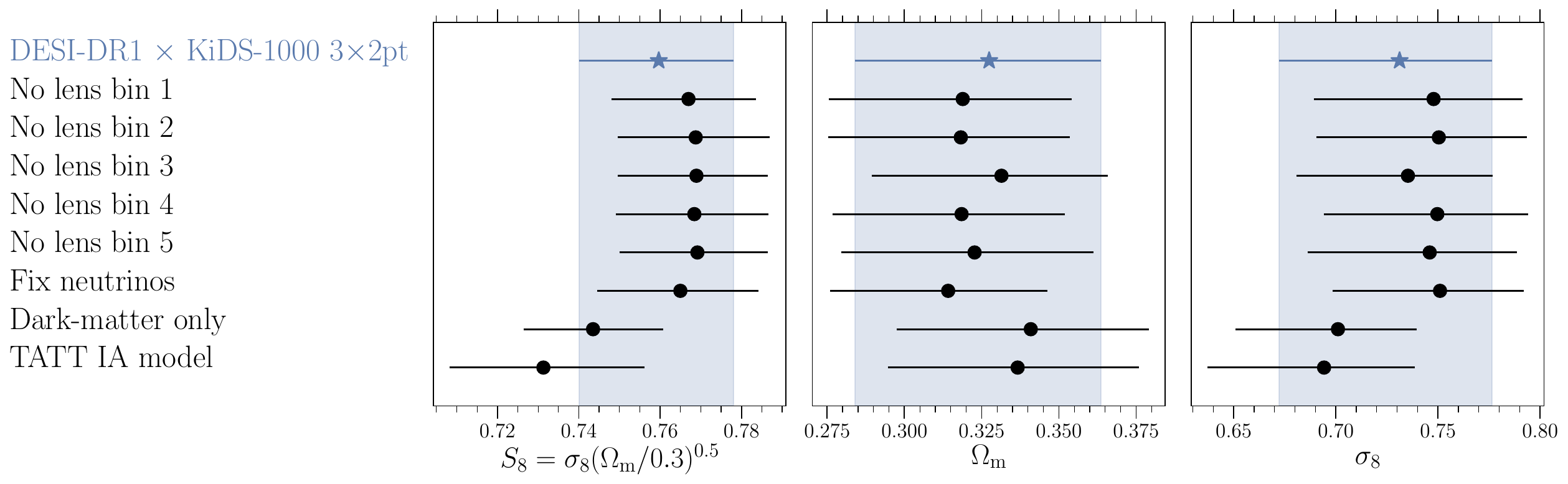}
    \includegraphics[width=0.9\linewidth]{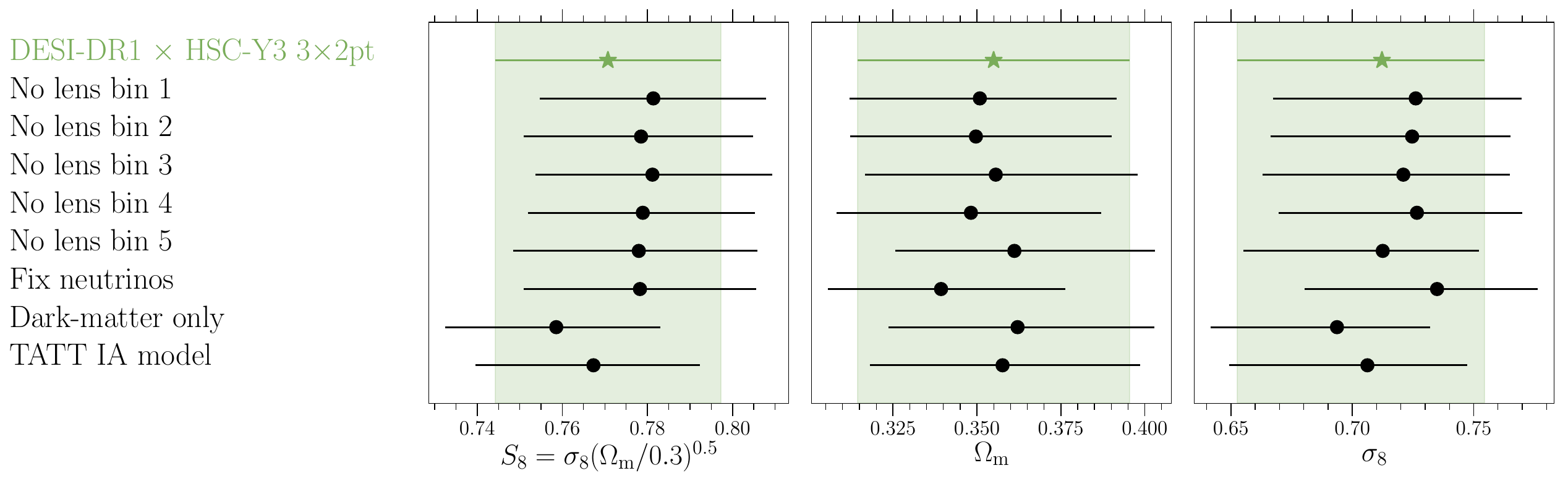}
    \caption{Comparison of 68\% C.I. constraints on $S_8$, $\Om$, and $\sigma_8$, in \lcdm\, from the  $3\times2$-pt combination of DESI-DR1 projected galaxy clustering and weak lensing data from DES-Y3, KiDS-1000, and HSC-Y3 for several variation of analysis choices.}
    \label{fig:robustness}
\end{figure*}

\subsection{Robustness tests}
\label{subsec:robustness}

In this section, we assess the robustness and internal consistency of our main results, i.e. the \threex\, cosmological constraints in \lcdm\, for each survey combination. We first re-analyse the data, removing one lens bin at a time, to test its internal consistency. In addition, we consider alternative modelling choices such as fixing the sum of the neutrino masses, using the TATT model for intrinsic alignments, and using the \textsc{HMcode2020} matter power spectrum without the impact of baryonic feedback effects (dark-matter only). The 1D marginalised constraints for each of these cases are shown in Fig.~\ref{fig:robustness}. 

Fig.~\ref{fig:robustness} demonstrates that our main results from \threex\, are robust to variations of the lens redshift bin and alternative modelling choices, with most of the shifts being much smaller than $1\sigma$. 
The largest shift corresponds to using the TATT model for intrinsic alignments, which shifts the $S_8$ constraint to lower values. This tendency has been observed in other analyses when switching from NLA to TATT (see, e.g., \citealt{2022PhRvD.105b3514A}, \citealt*{2022PhRvD.105b3515S}, \citealt{2023OJAp....6E..36D}, and \citealt{DECADE_cs_2025}) and is attributed to the extra freedom allowed by the TATT model, which enables lower $S_8$ values for some values of the TATT parameters. To assess whether the data prefers the TATT or the NLA model, we compute the ratio of Bayesian evidences for each survey combination. In Table~\ref{tab:tatt-constraints} we provide the $S_8$ constraints when assuming the TATT model along with the Bayesian ratio $R$ of TATT to NLA and goodness-of-fit statistics.
The ratio values indicate that TATT is disfavoured in the \threex\, analyses with DES-Y3 and HSC-Y3 data, but favoured in the case of KiDS-1000. Fig.~\ref{fig: comp IA TATT} shows that the DESI-DR1 $\times$ KiDS-1000 analysis is able to additionally constrain the $A_2$ and $\eta_2$ parameters, with $A_2=-1.83^{+0.57}_{-0.58}$, which explains the preference for the TATT model in this case (with $A_2$ being away from 0). The MAP estimate for $S_8$ in this case is 0.735, very similar to the mean of the posterior, indicating that the KiDS-1000 $S_8$ constraint with TATT is not impacted by prior volume effects\footnote{This is not the case for DES-Y3, in which we find a $\sim1\sigma$ difference between the mean and the MAP estimates. This shift could be due to prior volume effects from the extra unconstrained parameters $\eta_1$, $\eta_2$ and $b_{\mathrm{TA}}$ (see Fig.~\ref{fig: comp IA TATT}).}. Regarding goodness-of-fit metrics, we find very similar $p$-values when using TATT for the \threex\, combinations with DES-Y3 and HSC-Y3 compared to the fiducial values with NLA from Table~\ref{tab:lcdm-constraints}. In the case of KiDS-1000, we do find an improvement in the goodness-of-fit when using TATT, with $p=0.17$ instead of $0.083$, suggesting a preference for the TATT model.

Additionally, we analyse the \threex\, data vector with the dark-matter only version of \textsc{HMcode2020}, which is a less accurate model for the non-linear matter power spectrum, since we know that baryonic feedback effects are not negligible. We find some shifts to lower values of $S_8$, within the error bars, which demonstrates that our scale cuts (validated in \citealt{Nimas2025}) effectively mitigate the impact of mis-modelling baryonic feedback in the non-linear matter power spectrum. Another evidence of this is the fact that the posterior distribution of $\log_{10}(T_{\rm AGN / K})$ is prior dominated (see Fig.~\ref{fig: comp IA tagn}), since we have removed the scales most sensitive to this parameter.

For the Monte Carlo chains in which we remove one lens bin at a time, we use an emulator for \textsc{HMcode2020} developed by \cite{Tsedrik_2024} using \textsc{CosmoPower} \citep{SpurioMancini2022}. We note that the emulator has a slightly narrower range allowed for the parameters $\Om$, $\sum m_{\nu}$, and $\log T_{\rm agn}$ with respect to our fiducial priors, which introduces a slight shift to the constraints. We discuss and validate this comparison in Appendix~\ref{sec: hmemu}. 

\section{Comparison with previous analyses}
\label{sec:discussion}

\begin{figure}
    \centering
    \includegraphics[width=0.9\linewidth]{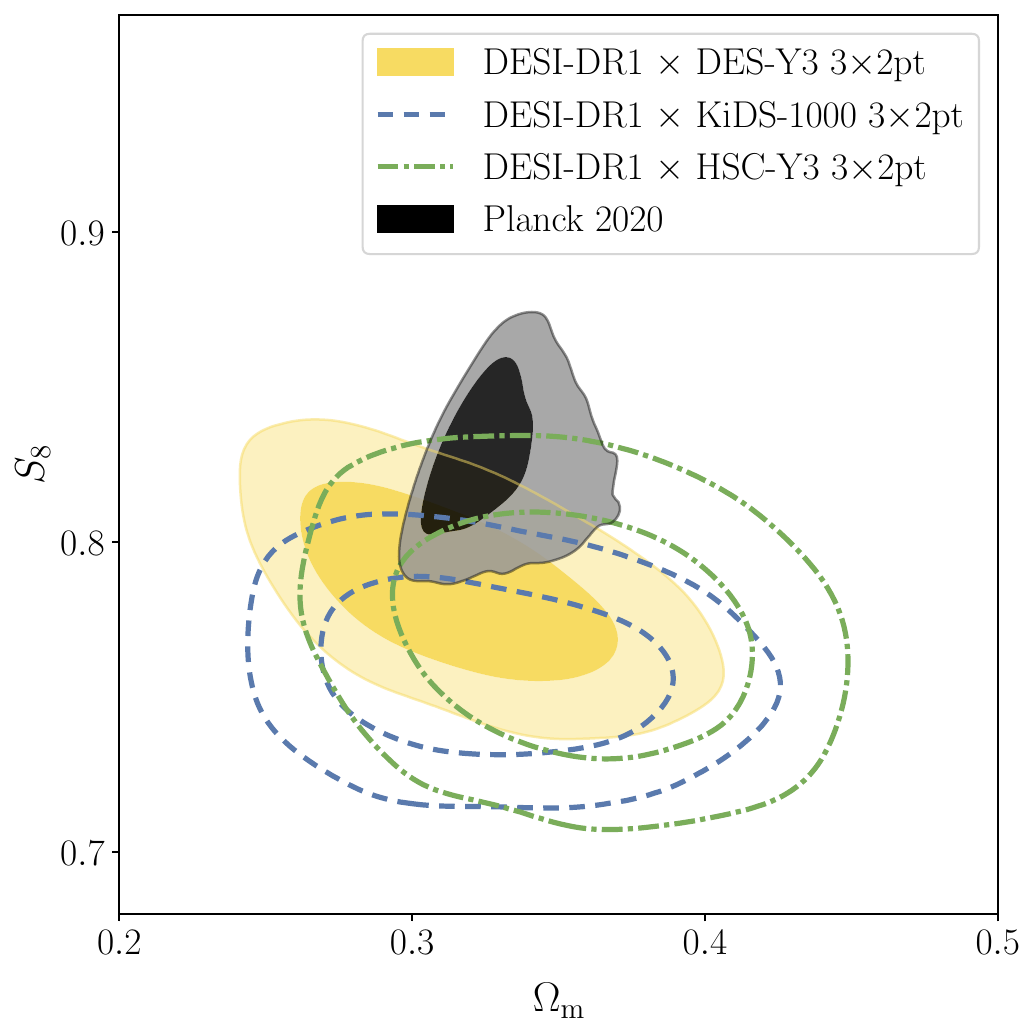}
    \caption{Marginalised constraints on $\Om$ and $S_8$ in \lcdm\, from the  $3\times2$-pt combination of DESI-DR1 projected galaxy clustering and weak lensing data from DES-Y3 (solid yellow), KiDS-1000 (dashed blue), and HSC-Y3 (dash-dotted green) compared to a re-analysis of the \cite{Planckcosmo} CMB primary anisotropies with a common set of cosmological parameters and priors (solid black).}
    \label{fig:planck_comp}
\end{figure}

\begin{figure}
    \centering
    \includegraphics[width=\linewidth]{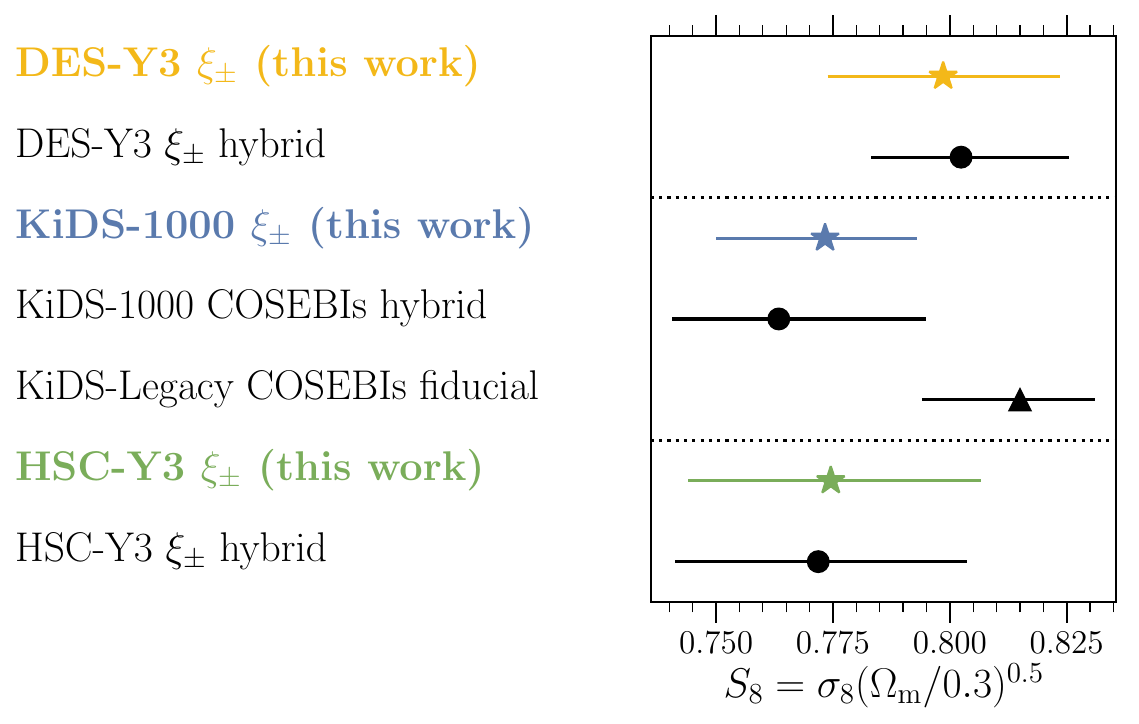}
    \caption{Marginalised constraints on $S_8$ in \lcdm\, from the DES-Y3, KiDS-1000, and HSC-Y3 $\xi_{\pm}$ measurements used in this work compared with those obtained with the ``\emph{hybrid}'' pipeline from \cite{2023OJAp....6E..36D}. Additionally, we compare with the recent KiDS-Legacy results from \cite{KiDSLegacy_CS_Wright2025}.}
    \label{fig: shear 1D comp}
\end{figure}

\begin{figure*}
    \centering
    \includegraphics[width=\linewidth]{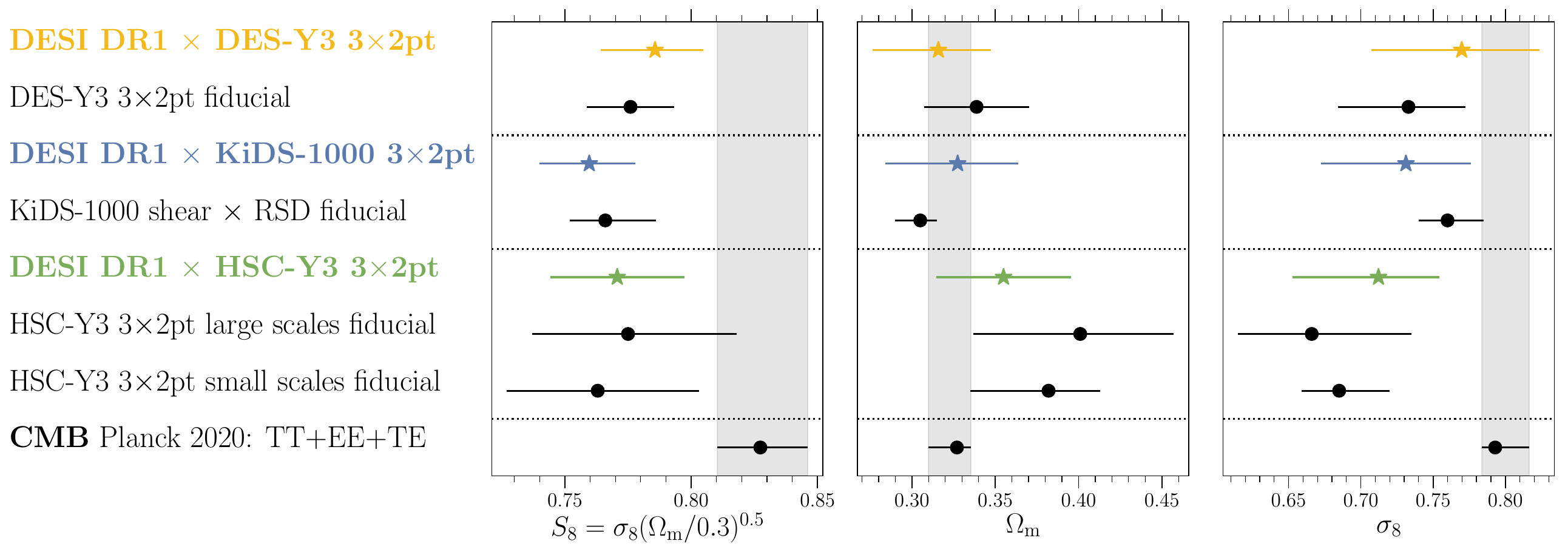}
    \caption{Marginalised constraints on $S_8$, $\Om$, and $\sigma_8$ in \lcdm\, from the $3\times2$-pt combination of DESI-DR1 projected galaxy clustering and weak lensing data from DES-Y3, KiDS-1000, and HSC-Y3 compared to the corresponding fiducial results from each collaboration: \cite{2022PhRvD.105b3520A} for DES-Y3, \cite{2021A&A...646A.140H} for KiDS-1000, and \cite{Sugiyama2023} and \cite{HSC3x2_Miyatake2023} for the HSC-Y3 large- and small-scale analyses, respectively. We also compare with a re-analysis of the \cite{Planckcosmo} CMB primary anisotropies with a common set of cosmological parameters and priors.}
    \label{fig:3x2 surveys comp}
\end{figure*}

In this section, we compare our main results in \lcdm\, with previous results from each WL survey and with primary anisotropies from the \cite{Planckcosmo} CMB observations, which have been reanalysed with a common set of priors. In particular, we include the primary temperature power spectra (TT) on scales $30 \le \ell \le 2508$, the E-mode and its cross power spectra with temperature (EE $+$ TE) in the range $30 \le \ell \le 1996$, and the low-$\ell$ temperature and polarisation likelihood (TT $+$ EE) at $2 \le \ell \le 29$. 

In Fig.~\ref{fig:planck_comp}, we compare the $S_8$ and $\Om$ constraints from each \threex\, survey combination with the CMB primary anisotropies from \cite{Planckcosmo}. We find that our combined DESI-DR1 constraints using the WL data from DES-Y3, KiDS-1000, and HSC-Y3 are consistent with each other. To quantify the consistency with CMB observations from \textit{Planck}, we estimate the parameter shift in the $S_8-\Om$ plane using the \textsc{tensiometer} package (see Sec.~\ref{subsec:gof} for details), finding the following shift significance values,
\begin{equation}
    \begin{gathered}
    \mathrm{DESI\text{-}DR1\, \times\, DES\text{-}Y3\quad vs.\quad Planck:}\qquad 1.3\sigma, \\
    \mathrm{DESI\text{-}DR1\, \times\, KiDS\text{-}1000\quad vs.\quad Planck:}\quad 2.1\sigma, \\
    \mathrm{DESI\text{-}DR1\, \times\, HSC\text{-}Y3\quad vs.\quad Planck:}\qquad  1.4\sigma.
    \end{gathered}
\end{equation}
We therefore do not conclude any statistical tension with early-Universe CMB observations, finding consistency at the $1.5-2\sigma$ level. While the cosmological constraints from each WL survey combination are consistent, a combined cosmic shear analysis would require a revalidation of the scale cuts at this increased precision (see, e.g., \citealt{2023OJAp....6E..36D}). We therefore leave such combination for future work.

In Fig.~\ref{fig: shear 1D comp}, we compare our reanalysis of the cosmic shear $\xi_{\pm}$ measurements from DES-Y3, KiDS-1000, and HSC-Y3 with the corresponding constraints presented in \cite{2023OJAp....6E..36D} using their ``\emph{hybrid}'' pipeline, given that it is the most similar to the modelling choices assumed in our analysis. We additionally compare with the recent KiDS-Legacy cosmic shear results from \cite{KiDSLegacy_CS_Wright2025}. For KiDS, we compare with their fiducial observable for cosmic shear, COSEBIs, instead of the real-space angular correlation functions $\xi_{\pm}$.  
The only differences between our modelling and inference pipeline and the \emph{hybrid} pipeline are: (1) sampling of $A_s$, $\Om$ and $\Omega_{\rm b}$ instead of $S_8$, $\omega_{\rm c}$, and $\omega_{\rm b}$, (2) higher prior edge for $\log_{10}(T_{\rm AGN}/K)$ (8.3 instead of 8), (3) lower prior edge for the sum of neutrino masses (starting at 0 instead of 0.055 eV), (4) use of the \textsc{Nautilus} sampler instead of \textsc{PolyChord} \citep{polychord}. 
We find consistent $S_8$ constraints when analysing the same data set, with only small shifts in the reported mean values that are compatible with sampling variance. In general, the 68\% credible intervals are slightly wider in our analysis due to the small change in priors, with the exception of KiDS-1000, for which we find smaller error bars than COSEBIs. The difference in constraining power in this case is due to having different scale sensitivities and a different degeneracy direction in the $\sigma_8-\Om$ plane. \cite{2021A&A...645A.104A} found a similar difference of constraining power between COSEBIs and $\xi_{\pm}$ when reporting $S_8$ constraints.
The fiducial cosmic shear constraints from KiDS-Legacy yield a higher value of $S_8$ (by $\sim2\sigma$) with respect to the KiDS-1000 constraints, bringing it closer to \textit{Planck}. This difference is discussed in Sec.~H.1 of \cite{KiDSLegacy_CS_Wright2025}, where they conclude that it is mainly due to changes to the redshift calibration sample and methodology, besides the inclusion of new area in KiDS-Legacy.

In Fig.~\ref{fig:3x2 surveys comp}, we compare our \threex\, results from the combination of DESI-DR1 projected galaxy clustering and weak lensing data from DES-Y3, KiDS-1000, and HSC-Y3 with the fiducial results of each collaboration using a similar combination of probes. For DES-Y3, we compare with their fiducial \threex\, analysis from \cite{2022PhRvD.105b3520A}, which uses angular correlation functions $\{ \xi_{\pm}(\theta)$, $\gamma_{\rm t}(\theta)$, $w(\theta) \}$ and includes information from shear ratios \citep*{2022PhRvD.105h3529S} in the likelihood inference. Besides the addition of information from shear ratios, DES is the only survey out of the three that has carried out so far a \threex\, analysis using only internal data. Their galaxy-galaxy lensing measurements, therefore, cover the whole DES footprint and have much higher signal-to-noise ratio than in the other analyses, in which galaxy-galaxy lensing is limited to the overlapping area between the spectroscopic and WL surveys. Given the small overlapping area between DESI-DR1 and DES-Y3 (see Table 1 and Fig.~1 of \citealt{2025arXiv250621677H}), we do not find an improvement in constraining power with respect to the fiducial DES-Y3 \threex\, analysis, even though we are using spectroscopic data for the projected galaxy clustering measurements.   

For KiDS-1000 we do not have a comparable analysis in terms of observables, since their fiducial result from \cite{2021A&A...646A.140H} uses redshift-space galaxy clustering from the Baryon Oscillation Spectroscopic Survey (BOSS), and additional spectroscopic data from the 2-degree Field Lensing Survey (2dFLenS) for the galaxy-galaxy lensing measurements. Their use of redshift-space galaxy clustering instead of projected galaxy clustering allows them to obtain much tighter constraints on $\Om$ and $\sigma_8$, given that they have access to the full 3D information from spectroscopic galaxy clustering. Thus, it is not a fair comparison with our analysis. We refer the reader to our companion paper that uses full-shape modelling of the DESI-DR1 anisotropic galaxy clustering \citep{Semenaite2025} for a comparison with KiDS-1000 on a more similar footing.

In the case of HSC-Y3 there are two fiducial \threex\, analyses, which use the Sloan Digital Sky Survey (SDSS) DR11 spectroscopic galaxy catalogue to measure projected galaxy clustering $w_{\rm p}$ and the galaxy-galaxy lensing observable $\Delta \Sigma$, i.e. the average excess surface mass density profile. \cite{Sugiyama2023} is the most similar to our analysis, since they also use a linear galaxy bias model and, therefore, restrict the analysis to the large scales. In particular, their scale cuts for galaxy-galaxy lensing and galaxy clustering are $(r_{\rm p,ggl},\, r_{\rm p,clus}) = (12,\, 8 )\,h^{-1}\,\mathrm{Mpc}$, while we include scales up to $6\,h^{-1}\,\mathrm{Mpc}$ for galaxy-galaxy lensing thanks to the point-mass marginalisation\footnote{We note that the linear galaxy bias scale cuts for the DES-Y1 \threex\, analysis, which did not include the point-mass marginalisation, were also $(r_{\rm p,ggl},\, r_{\rm p,clus}) = (12,\, 8 )\,h^{-1}\,\mathrm{Mpc}$ \citep{2018PhRvD..98d3526A}. }. \cite{HSC3x2_Miyatake2023}, instead, use the \texttt{Dark Emulator} \citep{DarkEmulator_Nishimichi2019} to model the $2\times2$-pt observables to quasi-nonlinear scales, which allows them to increase the constraining power of the analysis.

From Fig.~\ref{fig:3x2 surveys comp}, we see that our \threex\, 1D marginalised constraints on $S_8$, $\Om$, and $\sigma_8$ in \lcdm\, are consistent with the fiducial results from each WL survey, with differences smaller than 1$\sigma$. In this comparison, we were not expecting perfect consistency, given that we are using a new dataset for galaxy clustering from DESI-DR1. In addition to the potential statistical fluctuations from using new data, there are some differences in the modelling and analysis choices, scale cuts, and observables included, as summarised above.   

\section{Conclusions}
\label{sec:concl}

We have presented cosmological constraints from a joint analysis of DESI-DR1 projected galaxy clustering with weak lensing data from DES-Y3 \citep*{2022PhRvD.105b3514A,2022PhRvD.105b3515S}, KiDS-1000 \citep{2021A&A...645A.104A} and HSC-Y3 \citep{2022PASJ...74..421L}, under a unified modelling and analysis pipeline.

We first reanalyse the cosmic shear $\xi_{\pm}$ measurements from these three WL datasets, finding highly consistent determinations of $S_8$ with those presented in \cite{2023OJAp....6E..36D}, see Fig.~\ref{fig: shear 1D comp} for details.  We then obtain cosmological constraints from the $2\times2$-pt combination of DESI-DR1 projected galaxy clustering with galaxy-galaxy lensing measured in the overlapping footprint with each WL dataset. After ensuring consistency between cosmic shear and $2\times2$-pt results, we combine them to obtain our final \threex\, constraints from these datasets. 

In \lcdm, we measure $S_8= 0.786^{+0.022}_{-0.019}$ for DES-Y3 (2.6\% precision), $S_8= 0.760^{+0.020}_{-0.018}$ for KiDS-1000 (2.5\% precision), and $S_8= 0.771^{+0.026}_{-0.027}$ for HSC-Y3 (3.4\% precision). These results are consistent with each other, across different weak lensing surveys, and also compatible with CMB observations from \cite{Planckcosmo}. We quantify the consistency with \textit{Planck} at the $1.5-2\sigma$ level by means of parameter shift estimates in the $S_8-\Om$ plane. We have validated that these \threex\, results are robust to several data splits and analysis variations in Sec.~\ref{subsec:robustness}.

In $w$CDM, we measure the equation-of-state for dark energy to be $w= -1.12^{+0.32}_{-0.45}$ for the \threex\, combination with DES-Y3, $w=-0.88^{+0.18}_{-0.42}$ for KiDS-1000, and $w= -1.27^{+0.33}_{-0.40}$ for HSC-Y3. For all survey combinations, our results are consistent with the \lcdm\, value, $w=-1$. 

In Fig.~\ref{fig:3x2 surveys comp} we compare our main \lcdm\, \threex\, results with the corresponding fiducial ones from each WL survey, finding consistent constraints in general. Even though we are using spectroscopic galaxy clustering, and thus do not have to worry about photometric redshifts for the lenses, we still find some limitations in the constraining power due to: (1) the limited overlap between DESI-DR1 and the WL datasets, which lowers the signal-to-noise ratio of galaxy-galaxy lensing, (2) using projected galaxy clustering (2D) instead of anisotropic galaxy clustering measurements (3D information), (3) limitations in the modelling of the non-linear scales, which we do not include in our analysis.
Nevertheless, some companion papers provide an improvement regarding the last two points. \cite{Semenaite2025} and \cite{Joe_in_prep} will provide complementary cosmological constraints using DESI-DR1 redshift-space galaxy clustering measurements instead of projected galaxy clustering, while \cite{Lange2025} provides $2\times2$-pt cosmological constraints with a simulation-based approach that allows them to include smaller scales in the analysis.

In this work, we have used the cosmic shear measurements as provided by each WL collaboration, including their photometric redshift distributions and calibration. In the same way that we have unified the modelling and inference pipeline and the measurements of galaxy-galaxy lensing, one of the steps forward would be analysing these surveys under a unified photometric redshift calibration. Several companion papers have made advances in that regard: \cite{Lange2025_photoz} perform an independent, unified photometric redshift calibration of DES-Y3, KiDS-1000 and HSC-Y1 source catalogues using the new DESI COSMOS-XMM catalog from \cite{Ratajczak2025}, while \cite{Blanco2025} studies in more detail the potential for DESI to directly calibrate the KiDS-1000 source redshift distributions. In parallel, \cite{Ruggeri2025} provides estimates of the source redshift distributions of these WL datasets using clustering cross-correlation measurements with the DESI spectroscopic data. 

All this work sets the ground for future combined analyses of weak lensing data with DESI Data Release 2 (DR2) spectroscopic galaxy clustering, which includes the first three years of main survey observations. In addition to the greater data quality of DESI-DR2, we expect to have a larger overlapping area with the northern region of the DES footprint. At the same time, cross-correlating DESI spectroscopic galaxy clustering with other WL datasets in the north, such as the Dark Energy Camera All Data Everywhere (DECADE) \citep{DECADE_cat}, the Ultraviolet Near-Infrared Optical Northern Survey \citep{UNIONS}, and the \textit{Euclid} mission \citep{Euclidoverview_2025} in the future, will potentially increase the constraining power in a dramatic way. In this Stage-IV era of dark-energy surveys, it is critical to explore the synergies between different experiments and analyse them under a unified framework, to maximally exploit the wealth of information they provide, while assessing the robustness and consistency of different datasets.

\section*{Acknowledgments}

AP acknowledges financial support from the European Union's Marie Skłodowska-Curie grant agreement 101068581, and from the \textit{César Nombela} Research Talent Attraction grant from the Community of Madrid (Ref. 2023-T1/TEC-29011).  CB and NE acknowledge financial support from Australian Research Council Discovery Project DP220101609, and NE acknowledges financial support received through a Swinburne University Postgraduate Research Award.

This material is based upon work supported by the U.S. Department of Energy (DOE), Office of Science, Office of High-Energy Physics, under Contract No. DE–AC02–05CH11231, and by the National Energy Research Scientific Computing Center, a DOE Office of Science User Facility under the same contract. Additional support for DESI was provided by the U.S. National Science Foundation (NSF), Division of Astronomical Sciences under Contract No. AST-0950945 to the NSF’s National Optical-Infrared Astronomy Research Laboratory; the Science and Technology Facilities Council of the United Kingdom; the Gordon and Betty Moore Foundation; the Heising-Simons Foundation; the French Alternative Energies and Atomic Energy Commission (CEA); the National Council of Humanities, Science and Technology of Mexico (CONAHCYT); the Ministry of Science, Innovation and Universities of Spain (MICIU/AEI/10.13039/501100011033), and by the DESI Member Institutions: \url{https://www.desi.lbl.gov/collaborating-institutions}. Any opinions, findings, and conclusions or recommendations expressed in this material are those of the author(s) and do not necessarily reflect the views of the U. S. National Science Foundation, the U. S. Department of Energy, or any of the listed funding agencies.

The authors are honored to be permitted to conduct scientific research on I'oligam Du'ag (Kitt Peak), a mountain with particular significance to the Tohono O’odham Nation.

This work was supported through computational resources and services provided by the National Energy Research Scientific Computing Center (NERSC), a U.S. Department of Energy Office of Science User Facility operated under Contract No. DE-AC02-05CH11231. We also acknowledge the use of Spanish Supercomputing Network (RES) resources provided by the Barcelona Supercomputing Center (BSC) in MareNostrum 5 under allocations 2025-1-0045 and 2025-2-0046.

This research made use of the following python packages in addition to those already cited in the manuscript: {\sc astropy} \citep{astropy}, {\sc numpy} \citep{numpy}, {\sc scipy} \citep{scipy} and {\sc matplotlib} \citep{matplotlib}.

\newpage
\section*{Data availability}
Data points for the figures are available at \url{https://doi.org/10.5281/zenodo.17937581}. 

\bibliographystyle{mnras}

\bibliography{main}

\section*{Affiliations}
\scriptsize
\noindent
$^{1}$ CIEMAT, Avenida Complutense 40, E-28040 Madrid, Spain\\
$^{2}$ Institute for Astronomy, University of Edinburgh, Royal Observatory, Blackford Hill, Edinburgh EH9 3HJ, UK\\
$^{3}$ Ruhr University Bochum, Faculty of Physics and Astronomy, Astronomical Institute (AIRUB), German Centre for Cosmological Lensing, 44780 Bochum, Germany\\
$^{4}$ The Ohio State University, Columbus, 43210 OH, USA\\
$^{5}$ Centre for Astrophysics \& Supercomputing, Swinburne University of Technology, P.O. Box 218, Hawthorn, VIC 3122, Australia\\
$^{6}$ Department of Physics, American University, 4400 Massachusetts Avenue NW, Washington, DC 20016, USA\\
$^{7}$ Lawrence Berkeley National Laboratory, 1 Cyclotron Road, Berkeley, CA 94720, USA\\
$^{8}$ Department of Physics, Boston University, 590 Commonwealth Avenue, Boston, MA 02215 USA\\
$^{9}$ Department of Physics, The University of Texas at Dallas, 800 W. Campbell Rd., Richardson, TX 75080, USA\\
$^{10}$ Dipartimento di Fisica ``Aldo Pontremoli'', Universit\`a degli Studi di Milano, Via Celoria 16, I-20133 Milano, Italy\\
$^{11}$ INAF-Osservatorio Astronomico di Brera, Via Brera 28, 20122 Milano, Italy\\
$^{12}$ Department of Physics \& Astronomy, University College London, Gower Street, London, WC1E 6BT, UK\\
$^{13}$ Institut d'Estudis Espacials de Catalunya (IEEC), c/ Esteve Terradas 1, Edifici RDIT, Campus PMT-UPC, 08860 Castelldefels, Spain\\
$^{14}$ Institute of Space Sciences, ICE-CSIC, Campus UAB, Carrer de Can Magrans s/n, 08913 Bellaterra, Barcelona, Spain\\
$^{15}$ Instituto de Astrof\'{\i}sica de Canarias, C/ V\'{\i}a L\'{a}ctea, s/n, E-38205 La Laguna, Tenerife, Spain\\
$^{16}$ Department of Physics and Astronomy, The University of Utah, 115 South 1400 East, Salt Lake City, UT 84112, USA\\
$^{17}$ Instituto de F\'{\i}sica, Universidad Nacional Aut\'{o}noma de M\'{e}xico,  Circuito de la Investigaci\'{o}n Cient\'{\i}fica, Ciudad Universitaria, Cd. de M\'{e}xico  C.~P.~04510,  M\'{e}xico\\
$^{18}$ Department of Astronomy \& Astrophysics, University of Toronto, Toronto, ON M5S 3H4, Canada\\
$^{19}$ Department of Physics \& Astronomy and Pittsburgh Particle Physics, Astrophysics, and Cosmology Center (PITT PACC), University of Pittsburgh, 3941 O'Hara Street, Pittsburgh, PA 15260, USA\\
$^{20}$ Department of Physics, The Ohio State University, 191 West Woodruff Avenue, Columbus, OH 43210, USA\\
$^{21}$ University of California, Berkeley, 110 Sproul Hall \#5800 Berkeley, CA 94720, USA\\
$^{22}$ Institut de F\'{i}sica d’Altes Energies (IFAE), The Barcelona Institute of Science and Technology, Edifici Cn, Campus UAB, 08193, Bellaterra (Barcelona), Spain\\
$^{23}$ Departamento de F\'isica, Universidad de los Andes, Cra. 1 No. 18A-10, Edificio Ip, CP 111711, Bogot\'a, Colombia\\
$^{24}$ Observatorio Astron\'omico, Universidad de los Andes, Cra. 1 No. 18A-10, Edificio H, CP 111711 Bogot\'a, Colombia\\
$^{25}$ Center for Astrophysics $|$ Harvard \& Smithsonian, 60 Garden Street, Cambridge, MA 02138, USA\\
$^{26}$ NASA Einstein Fellow\\
$^{27}$ Institute of Cosmology and Gravitation, University of Portsmouth, Dennis Sciama Building, Portsmouth, PO1 3FX, UK\\
$^{28}$ University of Virginia, Department of Astronomy, Charlottesville, VA 22904, USA\\
$^{29}$ Fermi National Accelerator Laboratory, PO Box 500, Batavia, IL 60510, USA\\
$^{30}$ Institute of Astronomy, University of Cambridge, Madingley Road, Cambridge CB3 0HA, UK\\
$^{31}$ Institut d'Astrophysique de Paris. 98 bis boulevard Arago. 75014 Paris, France\\
$^{32}$ IRFU, CEA, Universit\'{e} Paris-Saclay, F-91191 Gif-sur-Yvette, France\\
$^{33}$ Department of Astronomy and Astrophysics, UCO/Lick Observatory, University of California, 1156 High Street, Santa Cruz, CA 95064, USA\\
$^{34}$ Center for Cosmology and AstroParticle Physics, The Ohio State University, 191 West Woodruff Avenue, Columbus, OH 43210, USA\\
$^{35}$ School of Mathematics and Physics, University of Queensland, Brisbane, QLD 4072, Australia\\
$^{36}$ Department of Physics, University of Michigan, 450 Church Street, Ann Arbor, MI 48109, USA\\
$^{37}$ University of Michigan, 500 S. State Street, Ann Arbor, MI 48109, USA\\
$^{38}$ NSF NOIRLab, 950 N. Cherry Ave., Tucson, AZ 85719, USA\\
$^{39}$ Department of Physics and Astronomy, University of California, Irvine, 92697, USA\\
$^{40}$ Department of Physics and Astronomy, University of Waterloo, 200 University Ave W, Waterloo, ON N2L 3G1, Canada\\
$^{41}$ Perimeter Institute for Theoretical Physics, 31 Caroline St. North, Waterloo, ON N2L 2Y5, Canada\\
$^{42}$ Waterloo Centre for Astrophysics, University of Waterloo, 200 University Ave W, Waterloo, ON N2L 3G1, Canada\\
$^{43}$ Sorbonne Universit\'{e}, CNRS/IN2P3, Laboratoire de Physique Nucl\'{e}aire et de Hautes Energies (LPNHE), FR-75005 Paris, France\\
$^{44}$ Department of Astronomy and Astrophysics, University of California, Santa Cruz, 1156 High Street, Santa Cruz, CA 95065, USA\\
$^{45}$ Departament de F\'{i}sica, Serra H\'{u}nter, Universitat Aut\`{o}noma de Barcelona, 08193 Bellaterra (Barcelona), Spain\\
$^{46}$ Instituci\'{o} Catalana de Recerca i Estudis Avan\c{c}ats, Passeig de Llu\'{\i}s Companys, 23, 08010 Barcelona, Spain\\
$^{47}$ Departamento de F\'{\i}sica, DCI-Campus Le\'{o}n, Universidad de Guanajuato, Loma del Bosque 103, Le\'{o}n, Guanajuato C.~P.~37150, M\'{e}xico\\
$^{48}$ Instituto Avanzado de Cosmolog\'{\i}a A.~C., San Marcos 11 - Atenas 202. Magdalena Contreras. Ciudad de M\'{e}xico C.~P.~10720, M\'{e}xico\\
$^{49}$ Space Sciences Laboratory, University of California, Berkeley, 7 Gauss Way, Berkeley, CA  94720, USA\\
$^{50}$ Instituto de Astrof\'{i}sica de Andaluc\'{i}a (CSIC), Glorieta de la Astronom\'{i}a, s/n, E-18008 Granada, Spain\\
$^{51}$ Departament de F\'isica, EEBE, Universitat Polit\`ecnica de Catalunya, c/Eduard Maristany 10, 08930 Barcelona, Spain\\
$^{52}$ Department of Physics and Astronomy, Sejong University, 209 Neungdong-ro, Gwangjin-gu, Seoul 05006, Republic of Korea\\
$^{53}$ Queensland University of Technology,  School of Chemistry \& Physics, George St, Brisbane 4001, Australia\\
$^{54}$ Max Planck Institute for Extraterrestrial Physics, Gie\ss enbachstra\ss e 1, 85748 Garching, Germany\\
$^{55}$ Department of Physics \& Astronomy, Ohio University, 139 University Terrace, Athens, OH 45701, USA\\
$^{56}$ National Astronomical Observatories, Chinese Academy of Sciences, A20 Datun Road, Chaoyang District, Beijing, 100101, P.~R.~China\\
\normalsize

\appendix
\section{HSC Year 1}
\label{sec: hsc-y1}

\begin{table*}
    \caption{HSC-Y1 constraints on $S_8$, $\Om$, and $\sigma_8$ in \lcdm\ and on $w$ in $w$CDM, with 68\% credible intervals using the mean 1D marginal posterior distribution. We report constraints on HSC-Y1 cosmic shear, the combination of DESI-DR1 projected galaxy clustering and galaxy-galaxy lensing ($2\times2$-pt), and the full $3\times2$-pt data vector. Along with the marginalised constraints, we provide goodness of fit statistics: the $\chi^2$ value at the best fit ($\chi^2_{\min}$), the total number of data points $N_{\rm D}$, the estimated effective number of free parameters $N_{\rm p, eff}$, the reduced $\chi^2_{\rm red}=\chi^2_{\min}/(N_{\rm D}-N_{\rm p, eff})$, and the goodness of fit probability $p$ (see Sec.~\ref{subsec:gof} for more details).  } 
    \label{tab:hscy1-constraints}
	\centering	
	\begin{tabular}{lccccccccc} 
		\toprule
		Analysis & $S_8$ & $\Om$ & $\sigma_8$ & $w$ & $\chi^2_{\min}$ & $N_{\rm D}$ & $N_{\rm p, eff}$ & $\chi^2_{\rm red}$ & $p(\chi^2_{\rm red})$\\
        \midrule
		\multicolumn{8}{l}{\textbf{\lcdm}} \\ [+0.1cm] 
        Cosmic shear & $0.858^{+0.028}_{-0.032}$ & $0.304^{+0.090}_{-0.041}$ &$0.872^{+0.110}_{-0.122}$ &  - & 201.52  & 170  & 4.59 & 1.22 & 0.029   \\ [+0.1cm] 
        $2\times2$-pt & $0.695^{+0.117}_{-0.109}$ & $0.343^{+0.048}_{-0.049}$ &$0.654^{+0.129}_{-0.106}$& - & 159.20 & 137  & 9.13  & 1.25 & 0.031   \\ [+0.1cm]
        $3\times2$-pt & $0.849^{+0.025}_{-0.027}$ & $0.318^{+0.049}_{-0.033}$ & $0.829^{+0.070}_{-0.059}$ & - & 362.64  &  307  & 10.62 &  1.22 & 0.005   \\
        \midrule

        \multicolumn{8}{l}{\textbf{$w$CDM}} \\ [+0.1cm] 
        $3\times2$-pt & $0.864^{+0.039}_{-0.048}$ & $0.325^{+0.048}_{-0.037}$ &$0.835^{+0.074}_{-0.064}$ & $-0.88^{+0.15}_{-0.43}$ & 361.35 & 307 & 10.77 & 1.22 & 0.006   \\

        \bottomrule
	\end{tabular}
\end{table*}

\begin{figure}
    \centering
    \includegraphics[width=\linewidth]{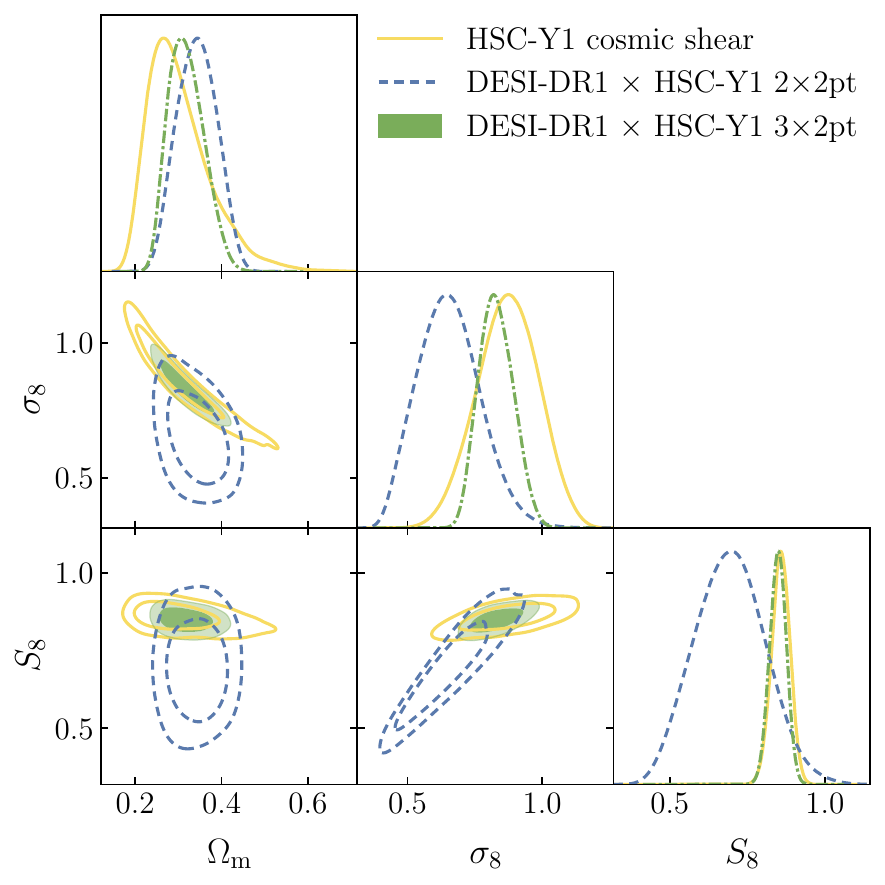}
    \caption{Marginalised constraints on $\Om$, $\sigma_8$, and $S_8$ in \lcdm\, for HSC-Y1 cosmic shear (solid yellow), the combination of galaxy-galaxy lensing and projected galaxy clustering from DESI-DR1 (dashed blue), and the $3\times2$-pt combination (solid green).}
    \label{fig:hscy1_lcdm}
\end{figure}

We have also carried out a \threex\, analysis using HSC Year 1 (Y1) data \citep{2018PASJ...70S..25M}, given that this survey combination was validated in the ``\emph{mock-challenge}'' paper by \cite{2025OJAp....8E..24B} and was also considered in the other supporting papers, such as \cite{2025arXiv250621677H} and \cite{Nimas2025}. In this case, the overlapping area with DESI-DR1 is 151.8~deg$^2$. We follow the fiducial HSC-Y1 cosmic shear analyses \citep{2019PASJ...71...43H,2020PASJ...72...16H} to set the Gaussian priors on the data-calibration nuisance parameters. For shear calibration, these priors are the same as those for HSC-Y3 in Table~\ref{tab: nuisance params}. Regarding photometric-redshift uncertainties, HSC-Y1 also assumes a Gaussian prior on the $\Delta z^j$ of each source bin $j$, centered at zero, and with $\sigma(\Delta z^j) = [\,0.0374, 0.0124, 0.0326, 0.0343\,]$. In this case, there are also two parameters that correct for PSF-related systematics, $\alpha_{\rm PSF}$ and $\beta_{\rm PSF}$. The corresponding Gaussian priors are  ($0.029, 0.01$) and ($-1.42, 1.11$), respectively, where we have expressed the Gaussian prior of the form $\mathcal{N}(\mu,\sigma)$.

In Table~\ref{tab:hscy1-constraints}, we report the 68\% credible intervals around the mean 1D marginal posterior for the main cosmological parameters (\,$S_8$, $\Om$, $\sigma_8$, and $w$\,), along with the goodness of fit metrics for each probe combination and cosmological model (\lcdm\, and $w$CDM). While the goodness-of-fit $p$ value for cosmic shear and $2\times2$-pt is above our threshold ($p>0.01$), this is not the case for the \threex\, combinations.
This is probably due to the high $S_8$ value preferred by the HSC Y1 $\xi_{+/-}$ cosmic shear constraints, which is considerably higher than the constraint from the 2$\times$2-pt part of the data vector. The fiducial HSC Y1 cosmic shear constraints in real space (see \citealt{2020PASJ...72...16H} after the erratum correction) are somewhat in tension with the harmonic space analysis from \cite{2019PASJ...71...43H} ($\sim1.6\sigma$ higher in $S_8$), which might indicate some unknown systematic with the data. We also note that the HSC Y1 footprint is small with many holes (see Fig.~1 from \citealt{2018PASJ...70S..25M}), which makes it more challenging to have an accurate analytical covariance matrix. 
For this reason, we report these constraints for completeness, but do not include them in the main results of this paper. In Fig.~\ref{fig:hscy1_lcdm}, we show the marginalised posterior distributions for $\Om$, $S_8$, and $\sigma_8$ in \lcdm. 

\section{Additional posterior distributions}
\label{sec:posteriors}

\begin{figure}
    \centering
    \includegraphics[width=\linewidth]{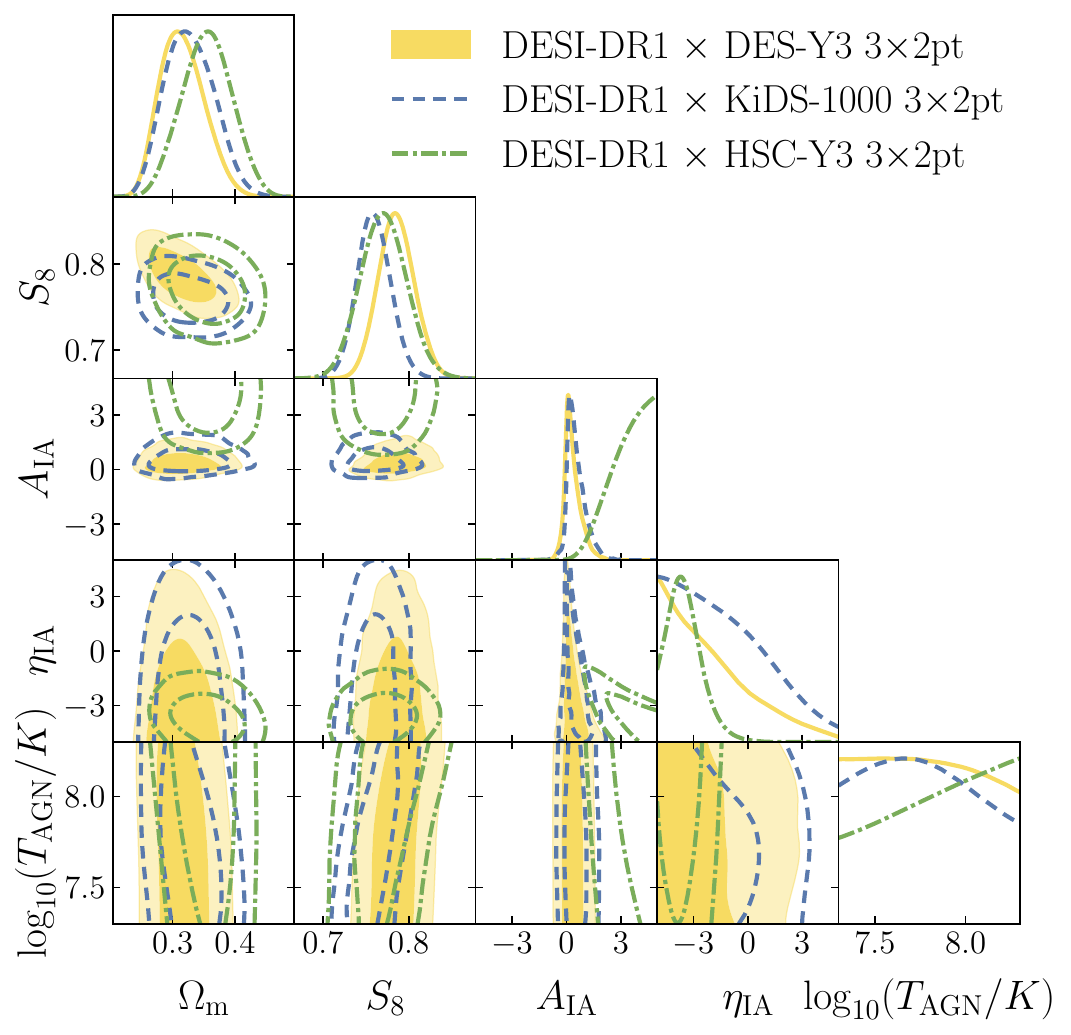}
    \caption{Marginalised constraints on $\Om$, $S_8$, and the IA and baryonic feedback parameters in \lcdm\, from the  $3\times2$-pt combination  of DESI-DR1 projected galaxy clustering and weak lensing data from DES-Y3 (solid yellow), KiDS-1000 (dashed blue), and HSC-Y3 (dash-dotted green).}
    \label{fig: comp IA tagn}
\end{figure}

\begin{figure}
    \centering
    \includegraphics[width=\linewidth]{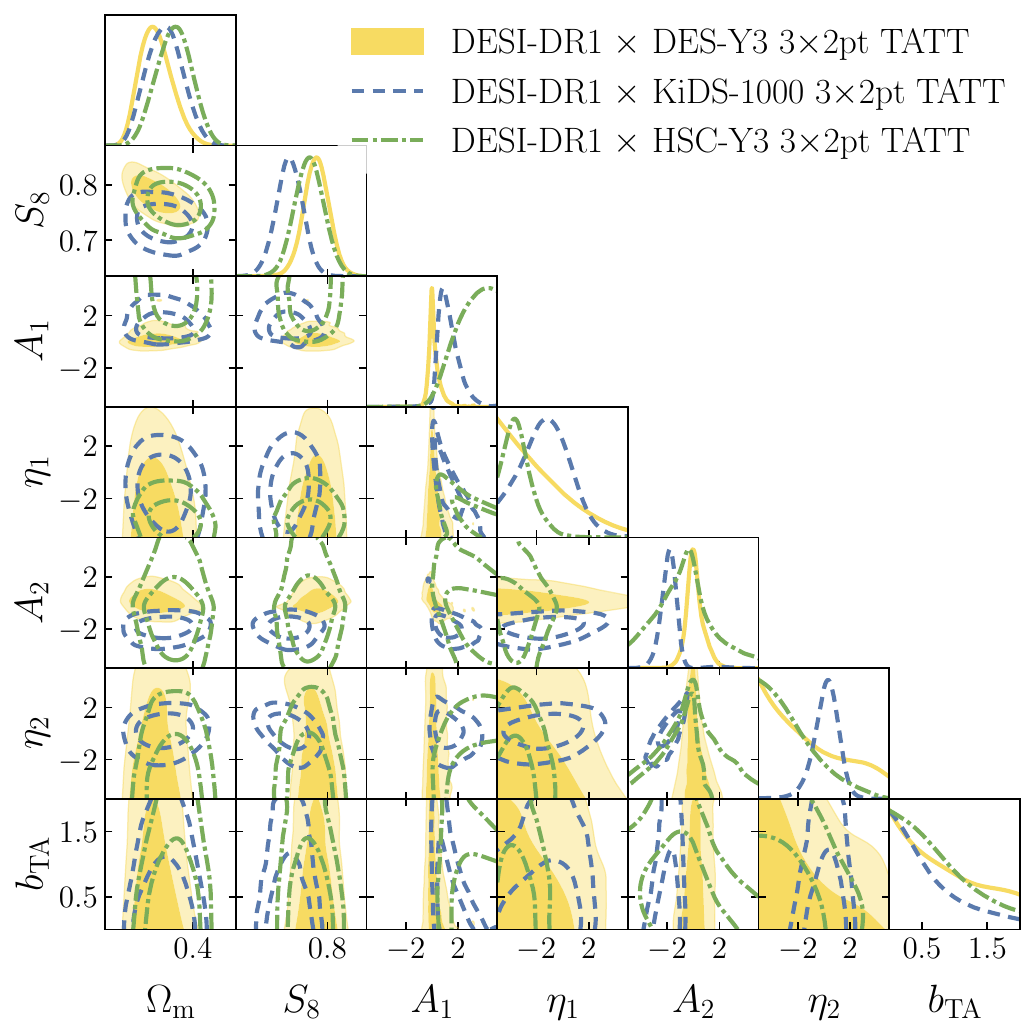}
    \caption{Marginalised constraints on $\Om$, $S_8$ and the IA parameters from the TATT model, in \lcdm\, from the  $3\times2$-pt combination  of DESI-DR1 projected galaxy clustering and weak lensing data from DES-Y3 (solid yellow), KiDS-1000 (dashed blue), and HSC-Y3 (dash-dotted green).}
    \label{fig: comp IA TATT}
\end{figure}

\begin{figure}
    \centering
    \includegraphics[width=\linewidth]{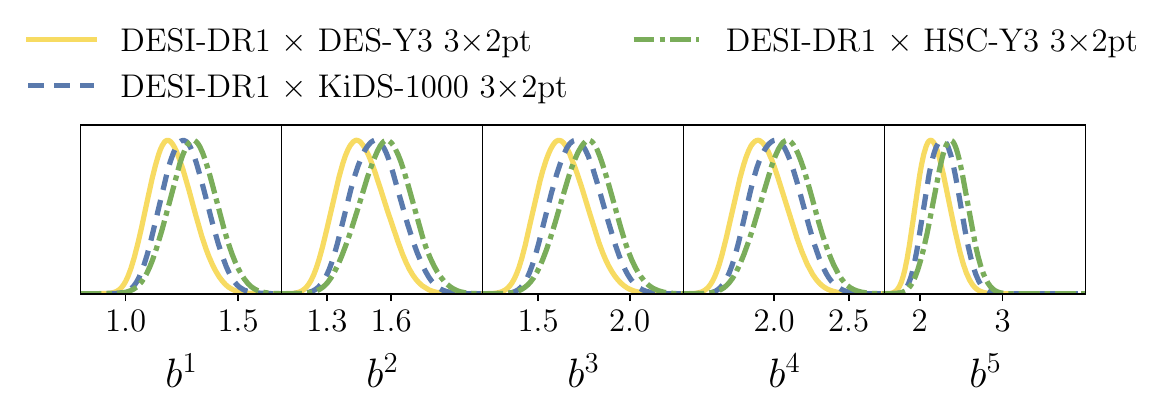}
    \includegraphics[width=\linewidth]{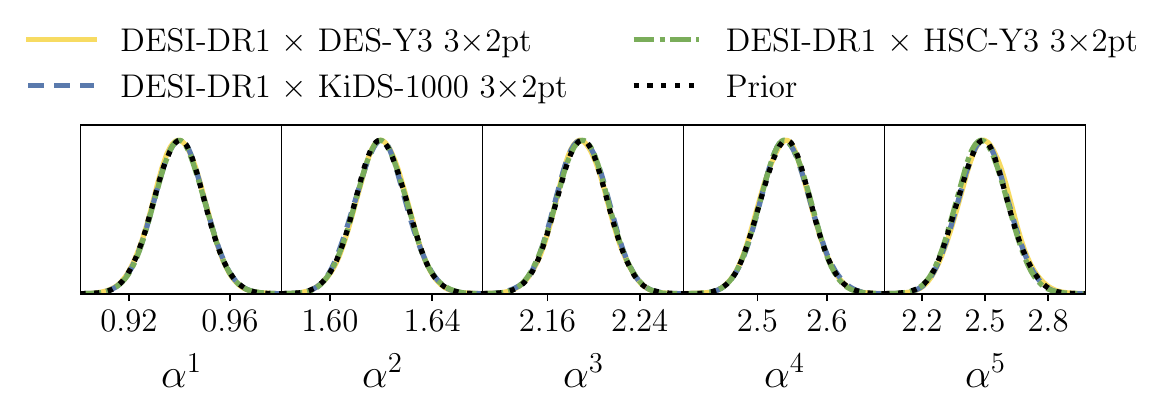}
    \caption{Marginalised constraints on the galaxy bias (top) and lens magnification (bottom) parameters, in \lcdm, from the  $3\times2$-pt combination of DESI-DR1 projected galaxy clustering and weak lensing data from DES-Y3 (solid yellow), KiDS-1000 (dashed blue), and HSC-Y3 (dash-dotted green). The priors assumed for the lens magnification parameters (see Table~\ref{tab:params}) are included in dotted black lines.}
    \label{fig: comp bias mag}
\end{figure}

\begin{figure*}
    \centering
    \includegraphics[width=0.8\linewidth]{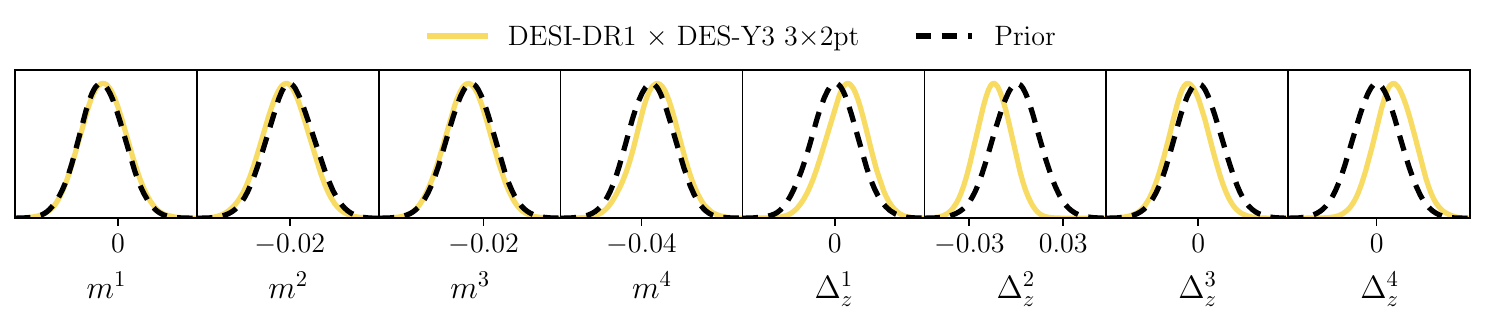}
    \includegraphics[width=0.8\linewidth]{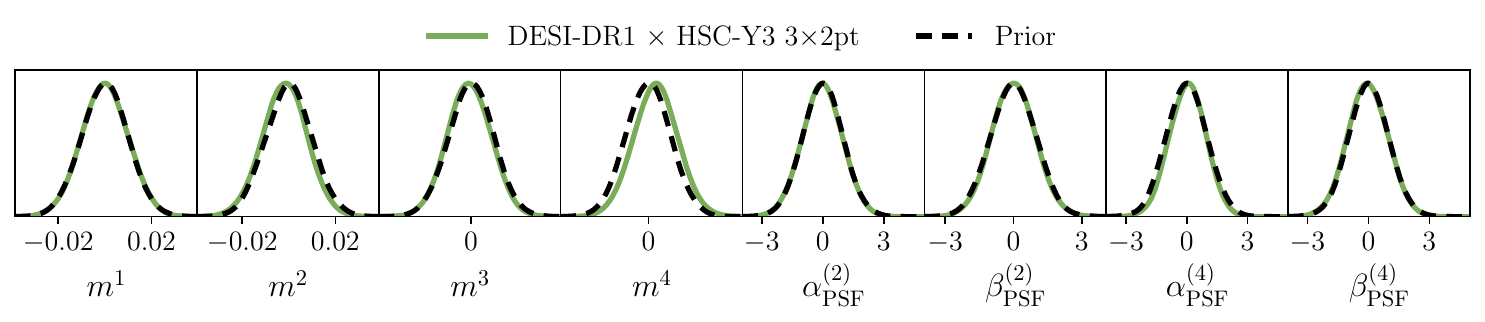}
    \includegraphics[width=0.6\linewidth]{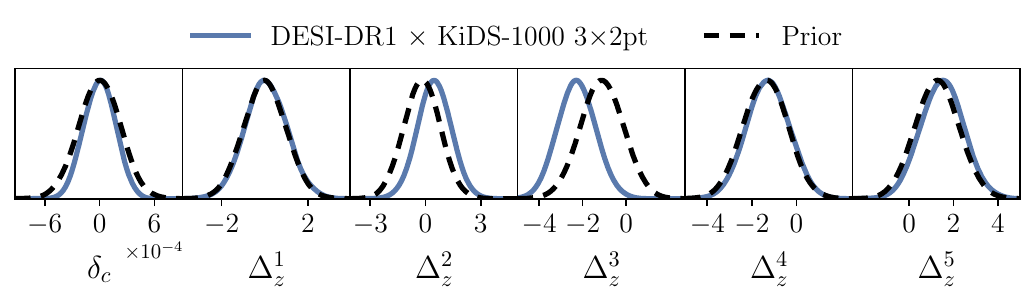}
    \caption{1D marginalised constraints on the shear nuisance parameters from the  $3\times2$-pt combination of DESI-DR1 projected galaxy clustering and weak lensing data from DES-Y3 (top), HSC-Y3 (middle), and KiDS-1000 (bottom). The priors assumed for these parameters (see Table~\ref{tab: nuisance params}) are included in dashed black lines.}
    \label{fig: nuisance}
\end{figure*}

\begin{figure}
    \centering
    \includegraphics[width=\linewidth]{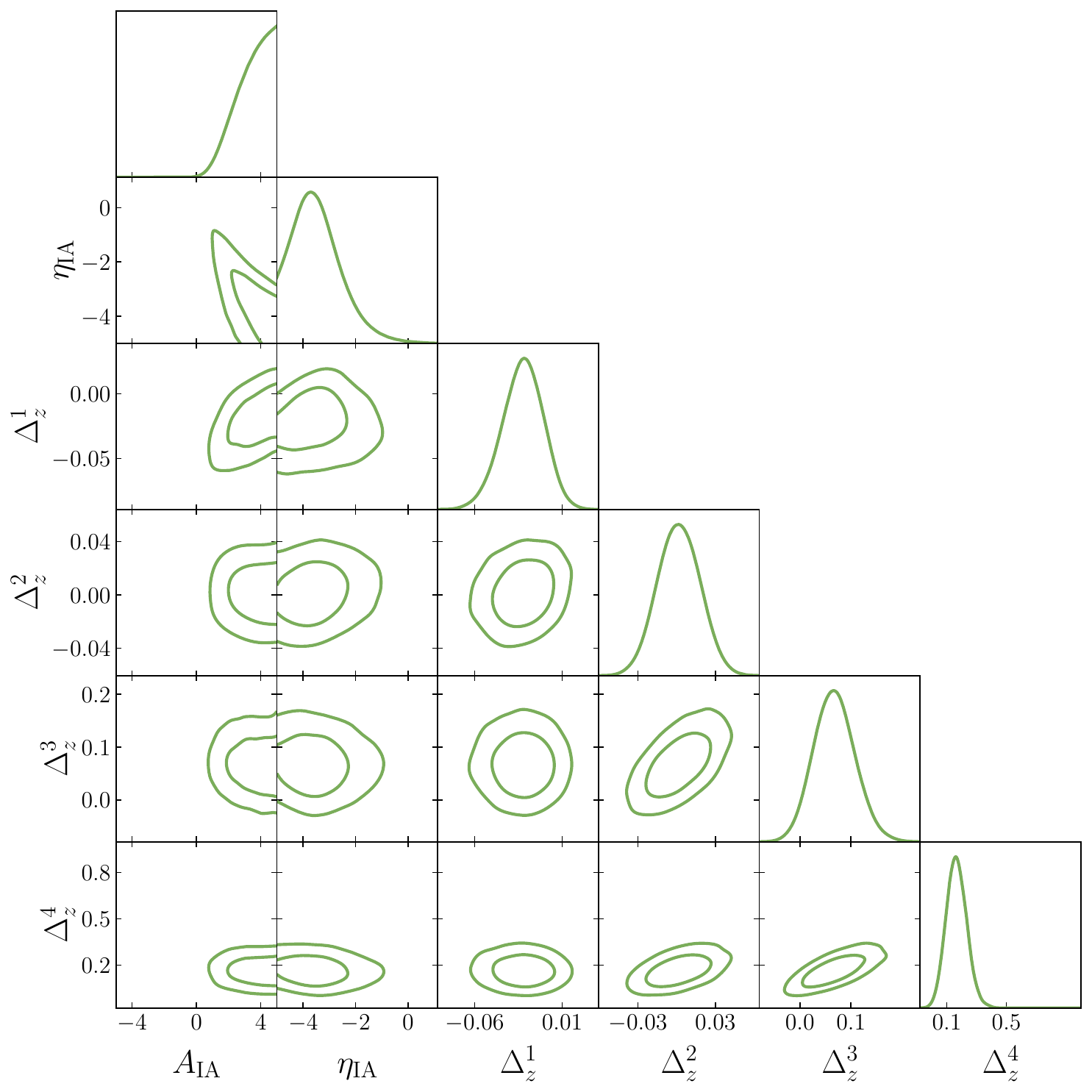}
    \caption{Marginalised constraints on the IA and $\Delta_z$ parameters, in \lcdm, from the  $3\times2$-pt combination of DESI-DR1 projected galaxy clustering and HSC-Y3 weak lensing data.}
    \label{fig: hscy3 IA delta z}
\end{figure}

In this Section, we include additional marginalised posterior distributions for the main \threex\, analyses in \lcdm\, presented in this work. Fig.~\ref{fig: comp IA tagn} shows the posterior distributions of $S_8$ and $\Om$, together with the IA NLA and baryonic feedback parameters. We find that the DESI-DR1 \threex\, combinations with DES Y3 and KiDS-1000 are able to constrain the amplitude of intrinsic alignments from our fiducial NLA model. However, the combination with HSC Y3 seems unable to constrain these parameters, which hit the edge of the priors. This is probably due to the wide flat priors assumed for $\Delta_z^3$ and $\Delta_z^4$, which the data has to self-calibrate. As a consequence, there is not enough constraining power to constrain the amplitude of intrinsic alignments $A_{\rm IA}$. The parameter $\log_{10}(T_{\rm AGN}/K)$ is in general unconstrained, since we have removed the scales most sensitive to baryonic feedback through the scale cuts validated in \cite{Nimas2025}. Similarly, in Fig.~\ref{fig: comp IA TATT}, we show the marginalised posterior distributions when using the TATT model for intrinsic alignments, where we find that the \threex\, combination with KiDS-1000 is able to constrain some of the additional IA parameters, like $A_2$ and $\eta_2$.   

In Fig.~\ref{fig: comp bias mag}, we compare the 1D marginalised posterior distributions on the galaxy bias and lens magnification parameters. The lens magnification parameters are not constrained by the data, and therefore we find that they agree perfectly with the Gaussian priors assumed (see Table~\ref{tab:params}). The galaxy bias posterior distributions, as expected, are consistent among the different survey combinations, since all the analyses share the same DESI-DR1 lenses.

In Fig.~\ref{fig: nuisance}, we present the 1D marginalised posterior distributions of most of the shear nuisance parameters assumed for DES-Y3, HSC-Y3, and KiDS-1000. We also overlay the Gaussian priors assumed for these parameters (see Table~\ref{tab: nuisance params}). We find that the data is able to constrain some of these parameters (especially the $\Delta_z$ ones). In all cases, we find consistency between these posterior distributions and the priors assumed (as required by our unblinding requirements).

In Fig.~\ref{fig: hscy3 IA delta z}, we show the marginalised posterior distributions of the IA and $\Delta_z$ parameters of the DESI-DR1 $\times$ HSC-Y3 \threex\, analysis in \lcdm. We do not find any degeneracy or strong correlation between the IA and photometric redshift parameters. Additionally, we see that the data is able to constrain the parameters $\Delta_z^3$ and $\Delta_z^4$, which have very wide flat priors instead of Gaussian ones (see Table~\ref{tab: nuisance params}).

\section{Validation of \texorpdfstring{\textsc{HMcode2020}}{HMcode2020} emulator}
\label{sec: hmemu}

\begin{figure}
    \centering
    \includegraphics[width=\linewidth]{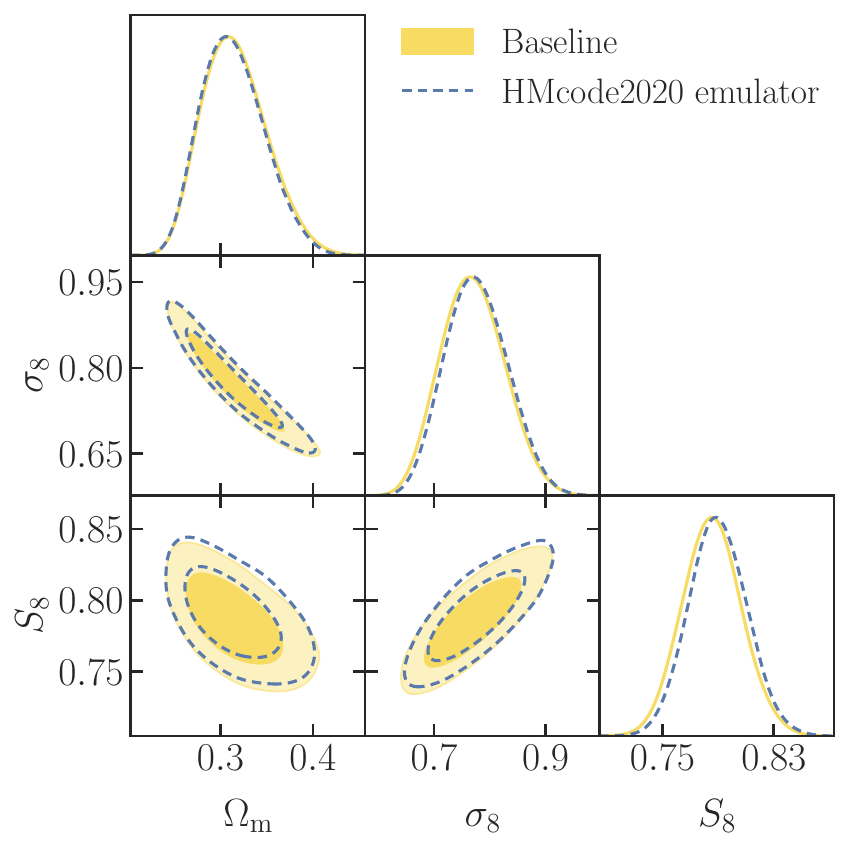}
    \caption{Comparison between the marginalised constraints on the parameters $\Om$, $\sigma_8$, and $S_8$ from the baseline DESI-DR1 $\times$ DES-Y3 $3\times2$-pt combined analysis (solid yellow) to a similar analysis using the \textsc{HMcode2020} emulator (dashed blue).}
    \label{fig:baseline_hmemu}
\end{figure}
As mentioned in Sec.~\ref{subsec:robustness}, we use the \textsc{HMcode2020} emulator developed by \cite{Tsedrik_2024} using \textsc{CosmoPower} \citep{SpurioMancini2022} to test the robustness of our results by excluding one lens redshift bin at a time. The parameter ranges of this emulator are narrower than those of our fiducial priors from Table~\ref{tab:params} on the $\Om$, $\Sigma\, m_{\nu}$, and $\log_{10}(T_{\rm AGN}/K)$ parameters. We therefore tighten the priors on these parameters to match the emulator's range in those cases, in the following way: 
(1) the higher edge prior of $\Om$ is reduced to 0.8 instead of 0.9, (2) the higher edge prior of $\Sigma\, m_{\nu}$ is reduced to 0.5 instead of 0.6, and (3) the lower edge prior for $\log_{10}(T_{\rm AGN}/K)$ is increased to 7.6 instead of 7.3. 

This change in priors slightly shifts the constraints on cosmological parameters. Fig.~\ref{fig:baseline_hmemu} shows the marginalised constraints on the $\Om$, $\sigma_8$, and $S_8$ cosmological parameters from the DES-Y3 $3\times2$-pt analysis using our fiducial pipeline  compared to the \textsc{HMcode2020} emulator. The difference in the 1D mean marginals in these parameters are $0.06\sigma$, $0.12\sigma$ and $0.2\sigma$, respectively. These shifts are small enough to enable the use of the emulator for the robustness tests in Sec.~\ref{subsec:robustness} and open the door to using it to accelerate future analyses.

\end{document}